\documentclass[a4paper,10pt]{article}
\usepackage{amsfonts}
\usepackage{amsmath}
\usepackage{amssymb}
\hoffset= - 1.0in         % switch off for draft style
\begin{document}
\begin{titlepage}
\hfill ITEP/TH-38/08

\vspace{5cm} \centerline{\begin{Huge}Correspondence between
Calogero-Moser systems\end{Huge}}
 \centerline{\begin{Huge}and
integrable SL(N,$\mathbb{C}$) Euler-Arnold tops\end{Huge}}
\vspace{1cm} \centerline{\begin{Large}Andrey Smirnov
\end{Large}} \vspace{1cm}
\centerline{\begin{Large}\texttt{E-mail:
asmirnov@itep.ru}\end{Large}} \vspace{5mm}
\centerline{\begin{Large} Moscow Institute of Physics and
Technology, Moscow, Russia \end{Large}} \vspace{5mm}
 \centerline{\begin{Large}
Institute for Theoretical and Experimental Physics,  Moscow,
Russia
\end{Large}}
\vspace{15mm} \centerline{\textbf{Abstract}} \vspace{3mm} The
equivalence between the N-particle Calogero-Moser systems and the
integrable sl(N,$\mathbb{C}$)-tops is shown. New rational and
trigonometric classical Lax operators for these systems are found.
Relations with new solutions of the classical Yang-Baxter
equations for sl(N,$\mathbb{C}$) are discussed. Explicit formulae
for N=2,3,4 are presented. \vspace{1cm}
\section{Introduction}
It was shown in \cite{Smir} that the systems of the
sl(2,$\mathbb{C}$) Euler-Arnold tops \cite{Arn,Olsh,Ves} are
equivalent to the two-particle Calogero-Moser systems (CM). In the
present paper we generalize these results to an arbitrary rank. We
show that the N-particle trigonometric and rational CM systems
correspond to some new classes of sl(N,$\mathbb{C}$) tops. The
L-operators for these systems do not coincide with the standard
degenerations of the elliptic Belavin L-operator. Thus, these Lax
operators correspond to some new solutions of the classical
Yang-Baxter equation for sl(N,$\mathbb{C}$). In appendix we
present explicit expressions for these L-operators for N=2,3,4. We
write out the integrals of motion and then check that they commute
with respect to the standard Poisson structure on
sl(N,$\mathbb{C}$).
\end{titlepage}
\section{Basic example in rank one}
The classical elliptic L-operator for sl(2,$\mathbb{C}$) is the
solution of the classical equation for the linear Poisson bracket:
\begin{equation}
\label{poiss}
 \{L^{ET}(z)\otimes1,1\otimes L^{ET}(z)\}=[r(z),1\otimes
L^{ET}(z)+L^{ET}(z)\otimes 1],
\end{equation}
where $r(z)$ is the classical elliptic Baxter r-matrix for
sl(2,$\mathbb{C}$). Explicitly for $L^{ET}(z)$ we get:
$$
L^{ET}(z)= \left[ \begin {array}{cc}
-\dfrac{\theta_{1\,1}^{\,\prime}(0)\,\theta_{1\,1}(z-\frac{1}{2})}{\theta_{1\,1}(z)\,\theta_{1\,1}(-\frac{1}{2})}\,S_{3}&
\dfrac{\theta_{1\,1}^{\,\prime}(0)\,e^{-\pi\,i\,z}}{\theta_{1\,1}(z)}\left(\dfrac{\theta_{1\,1}(z-\frac{\tau}{2})}{\theta_{1\,1}(-\frac{\tau}{2})}\,S_{1}
+\dfrac{\theta_{1\,1}(z-\frac{1+\tau}{2})}{\theta_{1\,1}(-\frac{1+\tau}{2})}\,S_{2}
\right)\\\dfrac{\theta_{1\,1}^{\,\prime}(0)\,e^{-\pi\,i\,z}}{\theta_{1\,1}(z)}\left(\dfrac{\theta_{1\,1}(z-\frac{\tau}{2})}{\theta_{1\,1}(-\frac{\tau}{2})}\,S_{1}
-\dfrac{\theta_{1\,1}(z-\frac{1+\tau}{2})}{\theta_{1\,1}(-\frac{1+\tau}{2})}\,S_{2}
\right) &
\dfrac{\theta_{1\,1}^{\,\prime}(0)\,\theta_{1\,1}(z-\frac{1}{2})}{\theta_{1\,1}(z)\,\theta_{1\,1}(-\frac{1}{2})}\,S_{3}
 \end {array} \right].
$$
Here $\theta_{k\,m}(z)$ are the standard theta functions with the
elliptic parameter $\tau$:
\begin{equation}
\label{tedef}
 \theta_{k\,m}(z,\,\tau)=\sum_{n\in\mathbb{Z}}\,\exp
\left( 2\,\pi\,i\,( (n+\frac{k}{2})^{2}\,\frac{\tau}{2}+
(z+\frac{m}{2})\,(n+\frac{k}{2}) ) \right).
\end{equation}
Equation (\ref{poiss}) fixes the Poisson brackets between $S_{i}$:
$$
\{\,S_{i}\,,\,S_{j}\,\}=2\,i\,\epsilon_{i\,j\,k}\,S_{k}.
$$
In what follows we also use the Chevalley basis:
$$
 S_{2}=i\,( S^{+}-S^{-} ), \ \ \ S_{1}=S^{+}+S^{-},\ \ \
\{S_{3}\,,\,S^{+}\}=2\,S^{+},  \ \ \ \{
S_{3}\,,\,S^{-}\}=-2\,S^{-}, \ \ \ \{S^{+}\,,\,S^{-}\}=S_{3}.
$$
In \cite{LOZ}, in the framework of Hitchin approach to the
integrable systems \cite{Hit}, the interrelations between the
elliptic sl(2,$\mathbb{C}$) top and the elliptic two particle
Calogero-Moser system was found. In terms of the Lax operators for
these systems the correspondence is looks like a gauge
transformation:
\begin{equation}
\label{eq1} L^{ET}(z)=\Xi(z)\,L^{ECM}(z)\,\Xi^{-1}(z),
\end{equation}
where $\Xi(z)$ is a 2$\times$2 matrix depending on the spectral
parameter $z$ and on the relative coordinate $u=u_{1}-u_{2}$ of
the CM system:
\begin{equation}
\Xi(z)=\Xi(z,u)=\left[ \begin {array}{cc} \theta_{0\,0}(z-2\,u ,
2\,\tau)&-\theta_{0\,0}(z+2\,u,2\,\tau)\\\noalign{\medskip}-\theta_{1\,0}(z-2\,u,2\,\tau)&\theta_{1\,0}(z+2\,u,2\,\tau)\end
{array}
 \right],
\end{equation}
and $L^{ECM}(z)$ is the L-operator for the elliptic CM system:
$$
L^{ECM}=\left[ \begin {array}{cc}
v&\nu\,\dfrac{\theta_{1\,1}(z+u)\,\theta_{1\,1}^{\,'}(0)}{\theta_{1\,1}(u)\,\theta_{1\,1}(z)}\\
\noalign{\medskip}\nu\,\dfrac{\theta_{1\,1}(z-u)\,\theta_{1\,1}^{\,'}(0)}{\theta_{1\,1}(-u)\,\theta_{1\,1}(z)}&-v\end
{array}
 \right].
$$
From (\ref{eq1}) we have:
\begin{equation}
\label{Ebos} \left\{
\begin{array}{l}
 S_{1} =
-\dfrac{\theta_{1\,0}(0)\theta_{1\,0}(2\,u)}{\theta_{1\,1}^{'}(0)\theta_{1\,1}(2\,u)}\,v+\dfrac{
\theta_{1\,0}^2(0)\theta_{0\,0}(2\,u)\theta_{0\,1}(2\,u)}{\theta_{0\,0}(0)
\theta_{0\,1}(0) \theta_{1\,1}^2(2\,u)}\nu,\\
\\
S_{2}=
\dfrac{\theta_{0\,0}(0)\theta_{0\,0}(2\,u)}{i\theta_{1\,1}^{'}(0)\theta_{1\,1}(2\,u)}\,v-\dfrac{
\theta_{0\,0}^2(0)\theta_{1\,0}(2\,u)\theta_{0\,1}(2\,u)}{i\theta_{1\,0}(0)
\theta_{0\,1}(0 )\theta_{1\,1}^2(2\,u)}\nu,\\
\\
S_{3}
=-\dfrac{\theta_{0\,1}(0)\theta_{0\,1}(2\,u)}{\theta_{1\,1}^{'}(0)\theta_{1\,1}(2\,u)}\,v+\dfrac
{\theta_{0\,1}^2(0)\theta_{0\,0}(2\,u)\theta_{1\,0}(2\,u)}{\theta_{0\,0}(0)\theta_{1\,0}(0
)\theta_{1\,1}^2(2\,u)}\nu.
\end{array}\right.
\end{equation}
Our general aim is to find the trigonometric and rational analogs
for these formulae. In the limit $\tau \rightarrow i\,\infty$
corresponding to the trigonometric degeneration expressions for
$S_{i}$ (\ref{Ebos}) diverge. To overcome this problem let us
apply to (\ref{eq1}) an additional gauge transformation
\cite{AHZ,GorskyZabrodin} depending on $q=\exp(\pi\,i\,\tau)$:
\begin{equation}
\label{gauge}
A^{T}(q)\,L^{ET}(z)\,A^{T}(q)^{-1}=A^{T}(q)\,\Xi(z)\,L^{EKM}(z)\,\Xi^{-1}(z)\,A^{T}(q)^{-1},
\end{equation}
where:
\begin{equation}
\label{GT} A^{T}(q)=\left[ \begin {array}{cc}
q^{\frac{1}{8}}&0\\\noalign{\medskip}0&q^{-\frac{1}{8}}\end
{array}
 \right].
 \end{equation}
 The gauge transformation acts in the
 "internal space" of the L-operator too, i.e. in addition to conjugation
 (\ref{gauge}) we should change the dynamical variables:
 $$
\left[ \begin {array}{cc} S_{3} &2\,S^{+}
\\\noalign{\medskip}2\,S^{-}&-S_{3}\end {array} \right]\rightarrow\left[ \begin {array}{cc}
q^{\frac{1}{8}}&0\\\noalign{\medskip}0&q^{-\frac{1}{8}}\end
{array}
 \right]^{-1}\, \left[ \begin {array}{cc} S_{3} &2\,S^{+}
\\\noalign{\medskip}2\,S^{-}&-S_{3}\end {array} \right]\,\left[ \begin {array}{cc}
q^{\frac{1}{8}}&0\\\noalign{\medskip}0&q^{-\frac{1}{8}}\end
{array}
 \right],
$$
Or more explicitly:
$$
S_{3}\rightarrow\,S_{3},\ \ \
S^{+}\rightarrow\,q^{-\frac{1}{4}}\,S^{+},\ \ \
S^{-}\rightarrow\,q^{\frac{1}{4}}\,S^{-}.
$$
After applying the gauge transformation of such a form, we can
take the limit $\tau\rightarrow i\infty$ in (\ref{gauge}). We get:
\begin{equation}
\label{Tbos} \left\{
\begin{array}{r}
S_{3}=-\dfrac{v}{u\,\tan(2\,\pi\,u)}-\dfrac{\nu}{\sin^{2}(2\,\pi\,u)},\\
\\
S^{+}=-\dfrac{v}{4\,\pi\,\sin(2\,\pi\,u)}-\dfrac{\nu\,\cos(2\,\pi\,u)}{4\,\sin^{2}(2\,\pi\,u)},\\
\\
S^{-}=\dfrac{v\,\cos^{2}(2\,\pi\,u)}{\pi\,\sin(2\,\pi\,u)}+\dfrac{\nu\,\cos(2\,\pi\,u)\left(1+sin^{2}(2\,\pi\,u)
\right)}{\sin^{2}(2\,\pi\,u)}.
\end{array}
\right.
\end{equation}
With respect to the canonical Poisson bracket $\{u,\,v\}=1$ we
have:
$$
\{S_{3}\,,\,S^{+}\}=2\,S^{+},  \ \ \ \{
S_{3}\,,\,S^{-}\}=-2\,S^{-}, \ \ \ \{S^{+}\,,\,S^{-}\}=S_{3}.
$$
The Casimir function:
$$
\Omega=S_{3}^2+4\,S^{+}\,S^{-}=\nu^2.
$$
Equations (\ref{Tbos}) can be presented in a form:
\begin{equation}
\label{eqt} L^{TT}(z)=\Xi^{T}(z)\,L^{TKM}(z)\,\Xi^{T}(z)^{-1},
\end{equation}
where:
$$
L^{TT}(z)=\lim_{q\rightarrow
0}A^{T}(q)\,L^{ET}(z)\,A^{T}(q)^{-1}\, ,\ \ \ \
\Xi^{T}(z)=q^{\frac{1}{8}}\,\lim_{q\rightarrow 0}
A^{T}(q)\,\Xi^{E}(z)\,,\ \
 L^{TCM}(z)=\lim_{q\rightarrow 0} L^{ECM}(z).
$$
Taking the limits we get:
$$
L^{TCM}(z)=\left[ \begin {array}{cc} v&\nu\,\pi( \cot(2\,\pi\,u)+
\cot(\pi\,z))\\\noalign{\medskip}\nu\,\pi( -\cot(2\,\pi\,u)+
\cot(\pi\,z))&-v\end {array}\right],
$$
$$
\Xi^{T}(z)=\left[ \begin {array}{cc}
1&-1\\\noalign{\medskip}-2\,\cos(\pi\,(z-2\,u))&2\,\cos(\pi(z+2\,u))\end
{array}\right],
$$
\begin{equation}
\label{trigLax} L^{TT}(z)=\left[ \begin {array}{cc}
\pi\,\cot(\pi\,z)\,S_{3}&\dfrac{2\,\pi\,S^{+}}{\sin(\pi\,z)}\\\noalign{\medskip}\dfrac{2\,\pi\,S^{-}}{\sin(\pi\,z)}+8\,\pi\,\,S^{+}\,sin(\pi\,z)&
-\pi\,\cot(\pi\,z)\,S_{3}\end {array} \right]\,.
\end{equation}
It is worth noting that L-operator (\ref{trigLax}) does not
coincide with the standard trigonometric degeneration of the
elliptical one. Thus, this Lax matrix corresponds to some new
trigonometric solution of the Yang-Baxter equation. Let us
calculate the hamiltonian for the system defined by $L^{TT}(z)$:
$$
{\rm Tr}\,(L^{TT}(z)^2)=2\,\
\pi^2\,(S_{3}^{2}-16\,S^{+2})-\dfrac{2\,\pi^2\,\Omega}{\sin^2(\pi\,z)},
$$
and from the definition for the hamiltonian we have:
\begin{equation}
H^{T}=\dfrac{1}{4\,\pi\,i}\oint \dfrac{{\rm tr}
{L^{TT}(z)^2}}{z}\,dz=\pi^2\,(S_{3}^{2}-16\,S^{+2}).
\end{equation}
Using (\ref{Tbos}) we see that this hamiltonian describes a motion
of the trigonometric CM system:
$$
H^{T}=\pi^2\,(S_{3}^{2}-16\,S^{+2})=v^2-\dfrac{\pi^2\,\nu^2}{\sin^{2}(2\,\pi\,u)}.
$$
Let us move to the rational case. To get the rational
degenerations of (\ref{Tbos}) we should replace the constant $\pi$
(the half period of the trigonometric functions) by a new variable
$x$ and then take the limit as $x$ goes to zero. Just as in the
trigonometric case we see that in this limit the right hand side
of (\ref{Tbos}) diverges. To avoid this difficulty we apply to
(\ref{eqt}) an additional gauge transformation:
\begin{equation}
\label{tt}
A^{R}(x)\,L^{TT}(z)\,A^{R}(x)^{-1}=A^{R}(x)\,\Xi^{T}(z)\,L^{TCM}(z)\,\Xi^{T}(z)^{-1}\,A^{R}(x)^{-1}\,,
\end{equation}
where the matrix $A^{R}(x)$ has the form:
\begin{equation}
\label{gaugeR} A^{R}(x)=\left[
\begin {array}{cc}
x&0\\\noalign{\medskip}\dfrac{2}{x}&\dfrac{1}{x}\end
{array}\right]\,.
\end{equation}
The gauge transformation in the internal space acts as follows:
$$
\left[ \begin {array}{cc} S_{3} &2\,S^{+}
\\\noalign{\medskip}2\,S^{-}&-S_{3}\end {array} \right]\longrightarrow \left[
\begin {array}{cc}
x&0\\\noalign{\medskip}\dfrac{2}{x}&\dfrac{1}{x}\end
{array}\right]^{-1}\, \left[ \begin {array}{cc} S_{3} &2\,S^{+}
\\\noalign{\medskip}2\,S^{-}&-S_{3}\end {array} \right]\,\left[
\begin {array}{cc}
x&0\\\noalign{\medskip}\dfrac{2}{x}&\dfrac{1}{x}\end
{array}\right],
$$
i.e. we have a substitution:
$$
S_{3}\rightarrow S_{3} + \dfrac{4\,S^{+}}{x^2}\,,\ \ \
S^{+}\rightarrow \dfrac{S^{+}}{x^2}\,,\ \ \ S^{-}\rightarrow
S^{-}\,x^2\,-2\,S_{3}\,-\dfrac{4\,S^{+}}{x^2}.
$$
After such a transformation we can take the limit $x=0$ in
(\ref{tt}) and we find:
\begin{equation}
\label{Rbos} \left\{
\begin{array}{l}
S_{3}=u\,v-\dfrac{\nu}{2},\\
\\ S^{+}=\dfrac{v}{2\,u}+\dfrac{\nu}{4\,u^2},\\
\\
S^{-}=-\dfrac{u^3\,v}{2}+\dfrac{3\,\nu\,u^2}{4}.
\end{array}
\right.
\end{equation}
With respect to the canonical Poisson bracket $\{ u,\,v\}=1$ we
have:
$$
\{S_{3}\,,\,S^{+}\}=2\,S^{+},  \ \ \ \{
S_{3}\,,\,S^{-}\}=-2\,S^{-}, \ \ \ \{S^{+}\,,\,S^{-}\}=S_{3},\ \ \
\Omega=S_{3}^2+4\,S^{+}\,S^{-}=\nu^2
$$
Expressions (\ref{Rbos}) can be presented in a form similar to
(\ref{eq1}):
\begin{equation}
\label{eqr} L^{RT}(z)=\Xi^{R}(z)\,L^{RCM}(z)\,\Xi^{R}(z)^{-1}\,,
\end{equation}
where
$$
 L^{RCM}(z)= \left[ \begin {array}{cc} v&\nu\left(
\dfrac{1}{z}+\dfrac{1}{2\,u}\right)\\\noalign{\medskip}\nu\left(
\dfrac{1}{z}-\dfrac{1}{2\,u}\right)&-v\end {array}\right]\,,
$$
$$
\Xi^{R}(z)= \left[ \begin {array}{cc}
-1&1\\\noalign{\medskip}-z\,u+u^2&-z\,u-u^2\end {array} \right]\,,
$$

\begin{equation}
\label{laxR}
 L^{RT}(z)= \left[ \begin {array}{cc}
\dfrac{S_{3}}{z}&\dfrac{2\,S^{+}}{z}\\\noalign{\medskip}\dfrac{2\,S^{-}}{z}+z\,S_{3}&-\dfrac{S_{3}}{z}\end
{array}
 \right]\,.
\end{equation}
Lax matrix (\ref{laxR}) is a new rational L-operator for
sl(2,$\mathbb{C}$) because it does not coincide with the standard
rational one. This  Lax matrix defines some new hamiltonian system
on the algebra sl(2,$\mathbb{C}$). The hamiltonian of this system
can be obtained from the expansion:
$$
{\rm tr}L^{RT}(z)^2= -4\,S^{+}\,S_{3}+\dfrac{2\,\Omega}{z^2}\ \
\Rightarrow H^{R}=2\,S^{+}\,S_{3}\,.
$$
Using (\ref{Rbos}) we see that this hamiltonian coincides with the
hamiltonian of the rational CM system:
\begin{equation}
H^{R}=2\,S^{+}\,S_{3}=v^2-\dfrac{\nu^2}{4\,u^2}\,.
\end{equation}
Thus, we have shown that there exist new trigonometrical and
rational L-operators corresponding to the two particle CM systems.
These new Lax operators can be obtained from the elliptical one by
applying to it some additional gauge transformations. One can
generalize this construction to an arbitrary rank. A main
difficulty here is to find the matrices of the additional gauge
transformations $A^{T}(q)$ and $A^{R}(x)$ for all N.
\section{The Elliptic SL(N) Top}
The elliptic sl(N,$\mathbb{C}$) top is the integrable system
defined by the classical Belavin-Drinfeld L-operator:
\begin{equation}
\label{ET} L^{ET}=\sum_{n+m\neq 0}^{N-1} S_{m n} \varphi_{m,n}(z)
T_{m n}\,.
\end{equation}
Here $\varphi_{m,n}(z)$ are defined by the standard theta
functions (\ref{tedef}):
\begin{equation}
\varphi_{m,n}(z)=\exp\left( -\dfrac{2 \, \pi \, i\, n\, z}{N}
\right) \phi \left( -\dfrac{m+n \, \tau}{N} ; z \right), \ \ \
\phi(u,z)=\dfrac{\theta_{1\,1}(u+z) \,
\theta_{1\,1}^{'}(0)}{\theta_{1\,1}(u)\, \theta_{1\,1}(z)}\,.
\end{equation}
where $S_{m\,n}$ are the generalized coordinates of the top and
$T_{m\,n}$ are the $N\times N$ matrices of the sl(N,$\mathbb{C}$)
generators in the fundamental representation. The generators
$T_{m\,n}$ correspond to the so called sin-alg basis and can be
constructed as follows:
$$
T_{m\,n}=\exp(\dfrac{\pi\,i\,m\,n}{N})\,Q^m\,\Lambda^n, \ \ \
(m=0...N-1,\, n=0...N-1,\, n+m\neq0)\,,
$$
$$
Q={\rm
diag}\left[\exp\left(\frac{2\,\pi\,i}{N}\right),\exp\left(\frac{4\,\pi\,i}{N}\right),...,\exp\left(\frac{2\,m\,\pi\,i}{N}\right),...1\right]\,,
$$
$$
\Lambda= \left[ \begin {array}{cccccc}
 0&1&0&..&..&0\\\noalign{\medskip}0&0&1&..&..
&0\\\noalign{\medskip}..&..&..&..&..&..\\\noalign{\medskip}..&..&..&
..&..&..
\\\noalign{\medskip}0&0&0&..&..&1\\\noalign{\medskip}1&0&0&..&..&0
\end {array} \right]\,.
$$
It is easy to see that these operators form the sl(N,$\mathbb{C}$)
basis, and obey the following relations:
\begin{equation}
[T_{s\,k}\,,\,T_{n\,j}]=2\,i\,\sin\left(
\dfrac{\pi}{N}\,(k\,n-s\,j)\right)\,T_{|s+n|, |k+j|},\ \ \ Tr(
T_{s\,k}\,T_{n\,j} )=\delta_{s\,,-n}\,\delta_{k\,,-j}\,N\,.
\end{equation}
Here $|a|$ is the value of $a$ modulo N that belongs to the
interval $(1..N)$, i.e.  $|0|=N$. The dynamics of the top is
defined by the Lax equation by means of the auxiliary M-matrix:
$$
\partial_{t}\,L^{E}(z)=[M^{E}(z)\,,L^{E}(z)]\,,\ \ \ \
M^{E}(z)=\sum_{n+m\neq
0}^{N-1}\exp\left(-\frac{n\,z}{N}\right)f\left(\frac{-m+n\,\tau}{N},z\right)\,S_{m\,n}\,T_{m\,n}\,,\
\ \ f(u,z)=\partial_{u}\phi(u,\,z)\,.
$$
From Lax equation we can get the equations of motion:
\begin{equation}
\partial_{t}\,S_{m\,n}=\dfrac{N}{\pi}\sum_{k\,l}
S_{k\,l}\,S_{m-k\,n-l}\,\wp\left( \dfrac{k+l\,\tau}{N},\,\tau
\right)\,\sin\,\dfrac{\pi}{N}(k\,n-m\,l)\,,
\end{equation}
where $\wp(z,\tau)$ is the Weierstrass elliptic function. The
quadratic hamiltonian of the system is defined by the expansion:
\begin{equation}
\dfrac{1}{N}\,{\rm
tr}L^{E}(z)^{2}=-2\,H^{E}+\Omega_{2}\,\wp(z,\,\tau), \ \ \
\Omega_{2}=\sum_{n+m\neq 0}^{N-1} S_{n\,m}^{2}\,,
\end{equation}
where
\begin{equation}
H^{E}=-\dfrac{1}{2}\,\sum_{m+n\neq0}^{N-1} \wp\left(
\dfrac{m+n\,\tau}{N}, \tau \right)\,S_{m\,n}\,S_{-m\,-n}\,.
\end{equation}
In \cite{LOZ} the correspondence between the elliptic
sl(N,$\mathbb{C}$) top and the N-particle elliptic CM system  was
established. The Lax operators of these systems are connected by
the equation:
\begin{equation}
\label{eqN} L^{ET}=\Xi(z)\,L^{ECM}(z)\,\Xi(z)^{-1}\,,
\end{equation}
where
$$
L^{EKM}(z)=v_{i}\,\delta_{i\,j}+(1-\delta_{i\,j})\,\nu\,\phi(u_{j}-u_{i},z),
$$
$$
\Xi(z)=\Xi^{'}(z) \times  {\rm diag} \left( (-1)^{l} \left(
\prod_{j<k , j,k \neq l} \theta_{1\,1}(u_{k}-u_{j} , \tau) \right)
^{-1} \right)\,,
$$
$$
\Xi^{'}(z)_{i\,j}=\theta_{\frac{i}{N}-\frac{1}{2} , \frac{N}{2}}
(z-N u_{j}, N \tau)\,.
$$
\section{The Trigonometric Lax Operator}
The trigonometric degeneration corresponds to the limit
$q\rightarrow 0$. It can be checked that in this limit the
right-hand side of (\ref{eqN}) diverges and as well as in the
sl(2,$\mathbb{C}$) case we need to apply to this equation some
additional regularising gauge transformation depending on $q$. To
find the matrix $A^{T}(q)$ of the additional gauge transformation
let us consider behaviour of the matrix elements of $\Xi(z)$ in
the neighbourhood of the point $q=0$. We find that:
\begin{equation}
\label{deg} \Xi_{i,j} \sim q^{\dfrac{N}{2} \left( \dfrac{i}{N} -
\dfrac{1}{2} \right)^2 },\ \ \ \Xi_{i,j}^{-1} \sim q^{ -
\dfrac{N}{2} \left( \dfrac{j}{N} - \dfrac{1}{2}  \right)^2  },\ \
\ q=\exp(2\, \pi \, i \tau ).
\end{equation}
From these expressions we find the leading terms for the right
hand side of (\ref{eqN}):
\begin{equation}
\label{T1}
 R_{i\,j}=\Xi_{i,k}(z) L_{k,s}^{CM}(z)
\Xi^{-1}_{sj}(z) \sim q^{\dfrac{N}{2} \left( \left(
\dfrac{i^2}{N^2}-\dfrac{i}{N} \right)   - \left(
\dfrac{j^2}{N^2}-\dfrac{j}{N} \right)   \right) }\,.
\end{equation}
Analogously to the sl(2,$\mathbb{C}$) case let us consider the
matrix $A^{T}(q)$ in the diagonal form:
\begin{equation}
\label{gaT}
 A^{T}(q)={\rm diag}\,[
\,q^{a_{1}},\,q^{a_{2}}...q^{a_{N}} ]\,.
\end{equation}
The degrees $a_{i}$ are defined from the condition for regularity
of (\ref{eqN}) after applying the the gauge transformation
$A^{T}(q)$ in the limit $q=0$. Combining (\ref{T1}) and
(\ref{gaT}) we have:
$$
\left[A^{T}(q)\,R\,A^{T}(q)^{-1}\right]_{i\,j}  \sim
q^{a_{i}-a_{j}+\dfrac{N}{2} \left( \left(
\dfrac{i^2}{N^2}-\dfrac{i}{N} \right)   - \left(
\dfrac{j^2}{N^2}-\dfrac{j}{N} \right)   \right) }\,.
$$
Therefore, the necessary condition  for $a_{i}$  has the following
form:
\begin{equation}
\label{c1} a_{j}-a_{i}=\dfrac{N}{2} \left( \left(
\dfrac{i^2}{N^2}-\dfrac{i}{N} \right)   - \left(
\dfrac{j^2}{N^2}-\dfrac{j}{N} \right)   \right)\,.
\end{equation}
The second condition is that the matrix $A^{T}(q)$ belongs to the
gauge group SL(N) for all $q$:
\begin{equation}
\label{c2} \det\,A^{T}(q)=1,\ \ \Rightarrow\ \ \sum_{i=1}^{N}
a_{i}=0\,.
\end{equation}
From (\ref{c1}) and (\ref{c2}) we get:
\begin{equation}
a_{i}=-\left(\dfrac{N}{2}\left( \dfrac{i^2}{N^2}-\dfrac{i}{N}
\right) + \dfrac{N^2-1}{12 N}\right)\,.
\end{equation}
After constructing the matrix $A^{T}(q)$ we can find the
trigonometric Lax operator:
\begin{equation}
\label{Tlim}
 L^{TT}(z)=\lim_{q\rightarrow0}\,A(q)\,L^{ET}(z)\,A^{-1}(q)
\end{equation}
Here as well as in the sl(2,$\mathbb{C}$) case in addition to
conjugation (\ref{Tlim}) the gauge transformation acts in the
internal space of the L-operator. To alleviate the calculations it
is useful to work in the basis $g_{i\,j}$ induced by the natural
embedding sl(N,$\mathbb{C}$) $\hookrightarrow$ gl(N,$\mathbb{C}$).
The connection between the sin-alg basis and gl(N,$\mathbb{C}$)
basis has the form:
$$
S_{n,m}=\sum_{|b-a|=n}  \exp\left(-\dfrac{\pi\,i\,m (|b-a|+2
|a|}{N} \right) \,g_{a\,b}\,,\ \ \ \
\{g_{i\,j},g_{k\,m}\}=\delta_{j\,k}\,g_{i\,m}-\delta_{i\,m}\,g_{k\,j}\,.
$$
To give the answer for the limit (\ref{Tlim}), which seems
complicated, it is useful to write it down by parts. For the
nondiagonal part we get:
\begin{equation}
\label{TLAX} \left\{
\begin{array}{ll}
 a<b & L^{TT}_{a,b}= \dfrac{N\,\pi}{\sin(\pi z)} \exp \left(-\dfrac{\pi\, i\, z}{N} (N-2b+2a)
 \right)\,g_{a,b}-\\ & 2 \,\pi\, i \,N\,\exp \left(\dfrac{2\pi i (b-a) z}{N} \right) \sum_{k=1}^{N-b} g_{a+k,b+k}\,,\\a>b,a
  \neq N & L^{TT}_{a,b}= \dfrac{N\,\pi}{\sin(\pi z)} \exp \left(-\dfrac{\pi\, i\, z}{N} (-N-2b+2a)
   \right)\,g_{a\,b}+\\ &  2 \,\pi\,N\, i \,\exp \left(-\dfrac{2\pi i (a-b) z}{N} \right) \sum_{k=1}^{b}
   g_{a-k,|b-k|}\,,
    \\a=N & L^{TT}_{N,b}=\dfrac{N\,\pi}{\sin(\pi z)} \exp\left(-\dfrac{\pi\,i\, (N-2\,b)\, z}{N}\right)g_{N\,b}-\\ &  2 \,\pi\, i \,N\,
    \exp \left(\dfrac{2\pi\, i \,b \,z}{N} \right)
      \sum_{k=1}^{N-b} g_{k,b+k}+\\ & 2 \, \pi\, i \,N\,
     \exp\left( \dfrac{2\, \pi \, i\,(b-N)\,z}{N}  \right)
     \sum_{k=0}^{b-1} g_{|N-b+k|,|N+k|}\,.
\end{array}
\right.,
\end{equation}
The diagonal part of $L^{TT}(z)$ is:
\begin{equation}
L_{a,a}^{TT}=\dfrac{\pi}{2\, \sin(\pi z)} \left[
A_{a}\exp(-\pi\,i\,z) + B_{a}\exp(\pi\,i\,z)   \right]\,,
\end{equation}
where:
\begin{equation}
A_{a}=\sum_{k=1}^{N}\left(\,  N-1-2|k-a|)\, \right)g_{k,k}\,,\ \ \
\ B_{a}=\sum_{k=1}^{N}\left(\, N-1-2|a-k|)\, \right)\,g_{k,k}\,.
\end{equation}
The connection with the N-particle trigonometric CM system has the
form:
\begin{equation}
\label{cont}
 L^{TT}(z)=\Xi^{T}(z)\,L^{TCM}(z)\,\Xi^{T}(z)^{-1}\,.
\end{equation}
where $L^{TCM}(z)$ is the Lax operator for the N-particle
trigonometric CM system:
\begin{equation}
L^{TCM}(u)_{i\,j}=v_{i}\,\delta_{i\,j}+(1-\delta_{i\,j})\,\nu\,\pi\,\left(\,\cot(\pi\,z)+\cot(\pi\,(u_{i}-u_{j}))\,\right)
\end{equation}
and the matrix $\Xi^{T}(z)$ is defined as the limit:
\begin{equation}
\Xi^{T}(z)=\lim_{q\rightarrow0\,} q^{\dfrac{N^2-1}{12
N}}\,A^{T}(q)\, \Xi(z) \,.
\end{equation}
Lax matrix (\ref{TLAX}) defines  some new integrable system on the
algebra sl(N,$\mathbb{C}$). The hamiltonian for this system is
defined in the standard way:
$$
H^{T}=\dfrac{1}{2\,\pi\,i}\oint \dfrac{{\rm
tr}L^{TT}(z)^2}{z}\,dz\,.
$$
Using equation (\ref{cont}) for the hamiltonian we get:
\begin{equation}
\label{corT} H^{T}=\dfrac{1}{2\,\pi\,i}\oint \dfrac{{\rm
tr}(L^{TT}(z)^2)}{z}\,dz=\dfrac{1}{2\,\pi\,i}\oint \dfrac{{\rm
tr}(L^{TCM}(z)^2)}{z}\,dz=\sum_{n=1}^{N}\,v^2_{n}+\sum_{i\neq
j}^{N}\dfrac{\pi^2\,\nu^2}{\sin^2\pi(u_{i}-u_{j})}=H^{TCM}\,.
\end{equation}
Therefore, we get the equivalence between the N-particle
trigonometric CM system and some new integrable top on the algebra
sl(N,$\mathbb{C}$) defined by L-operator (\ref{TLAX}). It worth
noting that this Lax has very sophisticated form, and it does not
coincide with the standard trigonometric L-operator for the
algebra sl(N,$\mathbb{C}$). To illustrate the formula (\ref{TLAX})
we give in appendix the explicit expressions for $L^{TT}(z)$ in
the cases N=2,3,4. We calculate the integrals of motion for the
system defined by these Lax and check that they commute.
\section{The Rational Case}
The rational degeneration corresponds to the limit as the period
of the trigonometric functions approaches infinity. To find this
degeneration we replace in (\ref{cont}) the constant $\pi$ by a
new variable $x$ and then tend $x$ to zero. As well as in the
sl(2,$\mathbb{C}$) case the right-hand side of (\ref{cont})
diverges in this limit. Analogously to the trigonometric case we
apply to this equation the regularizing gauge transformation
depending on the parameter $x$. Let us consider the matrix
$A^{R}(x)$ for this transformation in the form:
\begin{equation}
\label{gar} A^{R}(x)=W_{2}\, W_{1}\,,
\end{equation}
here the matrix $W{1}$ is lower-triangular and does not depend on
$x$. The matrix $W_{2}$ is diagonal:
\begin{equation}
W_{2}={\rm diag}(x^{b_{1}},x^{b_{2}},...,x^{b_{N}})\,.
\end{equation}
As in the previous section we need to find the degrees $b_{i}$ and
the coefficients of $W_{1}$  from the condition for regularity of
the limit:
$$
\lim_{x\rightarrow0}
A^{R}(x)\,\Xi^{T}(u)\,L^{TCM}(z)\,\Xi^{T}(u)^{-1}\,A^{R}(x)^{-1}\,.
$$
We find:
\begin{equation}
b_{i}=\left(-\frac{N(N-1)}{2\,N}+\frac{1-(i-1)N}{N}\right)\,
\end{equation}
\begin{equation}
\left\{
\begin{array}{ll}
W_{1\,i\,j}=\dfrac{(i-1)!}{(j-i)!\,(j-1)!},& i\neq N,
j\leq i\, ,\\
\\
W_{1\,N\,j}=\dfrac{N!}{i!\,(N-i)!}, & j\in(1\,...\,N)\,\\
\\
W_{1\,i\,j}=0,&j>i\,
\end{array}\right.
\end{equation}
Having found the regularizing matrix $A^{R}(x)$ we can get the
rational version of (\ref{cont}):
\begin{equation}
\label{eqr}
 L^{RT}(z)=\Xi^{R}(u)\,L^{RCM}(z)\,\Xi^{R}(u)^{-1}\,.
\end{equation}
where $L^{RCM}(z)$ is the Lax operator for the rational N-particle
CM system:
$$
L^{RCM}(u)_{i\,j}=v_{i}\,\delta_{i\,j}+\nu\,( 1-\delta_{i\,j}
)\,\dfrac{\nu}{u_{i}-u_{j}}\,,
$$
and the matrices $\Xi^{R}(z)$ and $L^{RT}(z)$ are defined as the
limits:
\begin{equation}
\label{lims} \Xi^{R}(z)=\lim_{x=0} x^{K_{N}}\,A^{R}(x)\, \Xi(z),\
\ \ \ L^{RT}(z)=\lim_{x=0} A^{R}(x)\,L^{TT}(z)\,A^{R}(x)^{-1}\,,
\end{equation}
where $K_{N}$ is some constant defined from the condition for
regularity of the limit for $\Xi^{R}(z)$. The hamiltonian of the
rational top is defined in the standard way:
$$
H^{R}=\dfrac{1}{2\,\pi\,i}\oint \dfrac{{\rm
tr}L^{RT}(z)^2}{z}\,dz\,.
$$
Using (\ref{eqr}) for this hamiltonian we get:
\begin{equation}
\label{corR} H^{R}=\dfrac{1}{2\,\pi\,i}\oint \dfrac{{\rm
tr}L^{RT}(z)^2}{z}\,dz=\dfrac{1}{2\,\pi\,i}\oint \dfrac{{\rm
tr}L^{RCM}(z)^2}{z}\,dz=\sum_{n=1}^{N}\,v^2_{n}+\sum_{i\neq
j}^{N}\dfrac{\nu^2}{(u_{i}-u_{j})^2}=H^{RCM}\,.
\end{equation}
The last equation gives the equivalence between the rational top
on the algebra sl(N,$\mathbb{C}$) defined by Lax operator
(\ref{eqr}) and the N-particle rational CM system. In this section
we have not presented the explicit expression for the $L^{RT}(z)$.
Nevertheless, using the matrix of additional gauge transformation
(\ref{gar}) by means of modern computational tools such as Maple
or Mathematica we can calculate limits (\ref{lims}) for N$\leq
10$. In appendix we present the examples of such a calculations
for N=2,3,4. We give the explicit expressions for the L-operators
corresponding to these cases and calculate the integrals of motion
for the rational tops and then check that they commute.
\\
\\
\begin{Large}\textbf{Acknowledgments} \end{Large}\\
\\
The author is grateful to M. Olshanetsky, A. Levin, A. Zotov and
A.Sleptsov for fruitful discussions and interest to this work. The
work was partly supported by RFBR grant 06-02-17382,   RFBR grant
06-01-92054-$KE_{a}$
and grant for support of scientific schools NSh-3036.2008.2.\\
\\
\\
\begin{Large}\textbf{Appendix T} \end{Large}\\
\\
Here we give the examples of the trigonometrical Lax operators for
N=2,3,4. We use the following normalization for Lax operators:
$$
L^{TT}(z)=\dfrac{\pi\,N}{\sin(\pi\,z)}\,L(z).
$$
\\
\textbf{The operator  L(z) for $N=2$}:\\
\\
\begin{small}
$$
\begin{array}{l}
 L=-1/2\,\cos \left( \pi \,z \right)  \left(
-g_{{1,1}}+g_{{2,2}}
\right)\\
\\
L_{1\,2}=g_{{1,2}}\\
\\
L_{2\,1}=-4\,g_{{1,2}} \left( \cos \left( \pi \,z \right)  \right)
^{2}+4\,g_{{1,2}}+g_{{2,1}}\\
\\
 L_{2\,2}=1/2\,\cos \left( \pi \,z \right)  \left( -g_{{1,1}}+g_{{2,2}} \right)
\end{array}
$$\\
\textbf{The operator L(z) for $N=3$}:\\
\\

$$
\begin{array}{l}
L_{1\,1}=(-{e^{-i\pi \,z}}g_{{3,3}}+{e^{-i\pi
\,z}}g_{{1,1}}-{e^{i\pi \,z}}g_{{2
,2}}+{e^{i\pi \,z}}g_{{1,1}})/3\\
\\
 L_{1\,2}={e^{-1/3\,i\pi \,z}}g_{{1,2}}-g_{{2,3}}{e^{5/3\,i\pi \,z}}+{e^{-1/3\,i
\pi \,z}}g_{{2,3}}\\
\\
L_{1\,3}=g_{{1,3}}{e^{1/3\,i\pi \,z}}\\
\\
L_{2\,1}=g_{{2,1}}{e^{1/3\,i\pi \,z}}+g_{{1,3}}{e^{1/3\,i\pi
\,z}}-{e^{-5/3\,i
\pi \,z}}g_{{1,3}}\\
\\
L_{2\,2}=({e^{-i\pi \,z}}g_{{2,2}}-{e^{-i\pi
\,z}}g_{{1,1}}-{e^{i\pi \,z}}g_{{3,
3}}+{e^{i\pi \,z}}g_{{2,2}})/3\\
\\
L_{2\,3}={e^{-1/3\,i\pi \,z}}g_{{2,3}}\\
\\
L_{3\,1}=-g_{{2,3}}{e^{5/3\,i\pi \,z}}+{e^{-1/3\,i\pi
\,z}}g_{{1,2}}-g_{{1,2}}{ e^{5/3\,i\pi \,z}}-{e^{-7/3\,i\pi
\,z}}g_{{2,3}}+g_{{3,1}}{e^{-1/3\,i \pi \,z}}+2\,{e^{-1/3\,i\pi
\,z}}g_{{2,3}}\\
\\
L_{3\,2}=g_{{3,2}}{e^{1/3\,i\pi \,z}}-g_{{1,3}}{e^{7/3\,i\pi
\,z}}+2\,g_{{1,3}}{e^{1/3\,i\pi \,z}}-{e^{-5/3\,i\pi
\,z}}g_{{2,1}}-{e^{-5/3\,i\pi
\,z}}g_{{1,3}}+g_{{2,1}}{e^{1/3\,i\pi \,z}}
\\
\\
L_{3\,3}=(-{e^{i\pi \,z}}g_{{1,1}}+{e^{-i\pi
\,z}}g_{{3,3}}-{e^{-i\pi \,z}}g_{{2,2}}+{e^{i\pi \,z}}g_{{3,3}})/3
\end{array}
$$
\\
\\
\textbf{The operator L(z) for $N=4$:}\\
\\
$$
\begin{array}{l}
L_{1\,1}=(-{e^{i\pi \,z}}g_{{3,3}}+3\,{e^{-i\pi
\,z}}g_{{1,1}}-3\,{e^{-i\pi \,z}}g_{{4,4}}+{e^{-i\pi
\,z}}g_{{2,2}}-3\,{e^{i\pi \,z}}g_{{2,2}}+3\,{e^{ i\pi
\,z}}g_{{1,1}}+{e^{i\pi \,z}}g_{{4,4}}-{e^{-i\pi
\,z}}g_{{3,3}})/8\\
\\
L_{1\,2}={e^{-1/2\,i\pi \,z}}g_{{3,4}}+{e^{-1/2\,i\pi
\,z}}g_{{1,2}}-g_{{3,4}}{e^{3/2\,i\pi \,z}}+{e^{-1/2\,i\pi
\,z}}g_{{2,3}}-g_{{2,3}}{e^{3/2\,i \pi \,z}}\\
\\
L_{1\,3}=g_{{2,4}}+g_{{1,3}}-g_{{2,4}}{e^{2\,i\pi \,z}}\\
\\
L_{1\,4}=g_{{1,4}}{e^{1/2\,i\pi \,z}}\\
\\
L_{2\,1}=g_{{1,4}}{e^{1/2\,i\pi \,z}}-{e^{-3/2\,i\pi
\,z}}g_{{1,4}}+g_{{2,1}}{e^{1/2\,i\pi \,z}}\\
\\
L_{2\,2}=(-3\,{e^{i\pi \,z}}g_{{3,3}}-3\,{e^{-i\pi
\,z}}g_{{1,1}}-{e^{-i\pi \,z}}g_{{4,4}}+3\,{e^{-i\pi
\,z}}g_{{2,2}}+3\,{e^{i\pi \,z}}g_{{2,2}}+{e^{i\pi
\,z}}g_{{1,1}}-{e^{i\pi \,z}}g_{{4,4}}+{e^{-i\pi
\,z}}g_{{3,3}})/8\\
\end{array}
$$
$$
\begin{array}{l}
\\
L_{2\,3}=-g_{{3,4}}{e^{3/2\,i\pi \,z}}+{e^{-1/2\,i\pi
\,z}}g_{{2,3}}+{e^{-1/2\,i\pi \,z}}g_{{3,4}}\\
\\
L_{2\,4}=g_{{2,4}}\\
\\
L_{3\,1}=g_{{2,4}}-{e^{-2\,i\pi \,z}}g_{{2,4}}+g_{{3,1}}\\
\\
L_{3\,2}=-{e^{-3/2\,i\pi \,z}}g_{{2,1}}+g_{{1,4}}{e^{1/2\,i\pi
\,z}}-{e^{-3/2\,i\pi \,z}}g_{{1,4}}+g_{{2,1}}{e^{1/2\,i\pi
\,z}}+g_{{3,2}}{e^{1/2\,i\pi \,z}}\\
\\
L_{3\,3}=(3\,{e^{i\pi \,z}}g_{{3,3}}-{e^{-i\pi
\,z}}g_{{1,1}}+{e^{-i\pi \,z}}g_{{4,4}}-3\,{e^{-i\pi
\,z}}g_{{2,2}}+{e^{i\pi \,z}}g_{{2,2}}-{e^{i\pi
\,z}}g_{{1,1}}-3\,{e^{i\pi \,z}}g_{{4,4}}+3\,{e^{-i\pi
\,z}}g_{{3,3}})/8\\
\\
L_{3\,4}={e^{-1/2\,i\pi \,z}}g_{{3,4}}\\
\\
L_{4\,1}=-g_{{2,3}}{e^{3/2\,i\pi \,z}}+2\,{e^{-1/2\,i\pi
\,z}}g_{{3,4}}+{e^{-1/2\,i\pi
\,z}}g_{{2,3}}+g_{{4,1}}{e^{-1/2\,i\pi \,z}}-{e^{-5/2\,i\pi
\,z}}g_{{3,4}}-g_{{1,2}}{e^{3/2\,i\pi \,z}}-g_{{3,4}}{e^{3/2\,i\pi
\,z}}+{e^{-1/2\,i\pi \,z}}g_{{1,2}}\\
\\
L_{4\,2}=-{e^{-2\,i\pi
\,z}}g_{{2,4}}+2\,g_{{2,4}}+g_{{3,1}}+g_{{1,3}}-g_{{2,4}
}{e^{2\,i\pi \,z}}-g_{{1,3}}{e^{2\,i\pi
\,z}}+g_{{4,2}}-{e^{-2\,i\pi\,z}}g_{{3,1}}\\
\\
L_{4\,3}=2\,g_{{1,4}}{e^{1/2\,i\pi \,z}}-{e^{-3/2\,i\pi
\,z}}g_{{1,4}}-g_{{1,4}}{e^{5/2\,i\pi \,z}}-{e^{-3/2\,i\pi
\,z}}g_{{2,1}}+g_{{4,3}}{e^{1/2\,i\pi \,z}}+g_{{2,1}}{e^{1/2\,i\pi
\,z}}-{e^{-3/2\,i\pi \,z}}g_{{3,2}}+g_{{3,2}}{e^{1/2\,i\pi \,z}}\\
\\
L_{4\,4}=({e^{i\pi \,z}}g_{{3,3}}+{e^{-i\pi
\,z}}g_{{1,1}}+3\,{e^{-i\pi \,z}}g_{{4,4}}-{e^{-i\pi
\,z}}g_{{2,2}}-{e^{i\pi \,z}}g_{{2,2}}-3\,{e^{i\pi
\,z}}g_{{1,1}}+3\,{e^{i\pi \,z}}g_{{4,4}}-3\,{e^{-i\pi
\,z}}g_{{3,3}})/8
\end{array}
$$
\\
\\
\textbf{The hamiltonians and highest integrals of motion}\\
\\
On calculating the integrals of motion we use the standard
definition:
$$
H_{k}=\dfrac{1}{2\,k\,\pi\,i}\oint\,\dfrac{{\rm tr}
\left(L^{TT}(z)\right)^{k}}{z}\,dz.
$$
The integral $H_{2}$ is usually identified with the hamiltonian.
We check the commutativity with respect to the standard Poisson
bracket on the algebra gl(N,$\mathbb{C}$):
$$
\{A\,,\,B\}_{N}=\sum_{i\,j\,k\,=1}^{N}\,\left(\dfrac{\partial\,A}{\partial\,g_{i,j}}\,\dfrac{\partial\,B}{\partial\,g_{j,k}}\,-
\dfrac{\partial\,B}{\partial\,g_{i,j}}\,\dfrac{\partial\,A}{\partial\,g_{j,k}}\right)\,g_{i\,k}
$$\\
\\
 \textbf{The case N=2}\\

$$
 H_{2}=1/2\, \left( g_{{2,2}}-g_{{1,1}}+4\,g_{{1,2}} \right)
\left( g_{{2,2} }-g_{{1,1}}-4\,g_{{1,2}} \right) {\pi }^{2}
$$
\textbf{The case N=3}
\\
$$
\begin{array}{l}
H_{2}=2\, \left(
{g_{{2,2}}}^{2}-g_{{2,2}}g_{{3,3}}-9\,g_{{2,3}}g_{{2,1}}-g_
{{1,1}}g_{{3,3}}-9\,g_{{1,2}}g_{{1,3}}+{g_{{1,1}}}^{2}-27\,g_{{1,3}}g_
{{2,3}}-g_{{1,1}}g_{{2,2}}+{g_{{3,3}}}^{2} \right) {\pi }^{2}\\
\\
H_{3}= {g_{{1,1}}}^{2}g_{{2,2}}+27\,{g_{{1,3}}}^{2}g_{{2,1}}-{g_{
{1,1}}}^{2}g_{{3,3}}+{g_{{2,2}}}^{2}g_{{3,3}}-9\,g_{{1,2}}g_{{2,2}}g_{
{1,3}}-27\,{g_{{2,3}}}^{3}+9\,g_{{2,3}}g_{{2,1}}g_{{1,1}} \\
-g_{{1,1}}{g_
{{2,2}}}^{2}+g_{{1,1}}{g_{{3,3}}}^{2}+27\,{g_{{1,3}}}^{3}-27\,{g_{{2,3
}}}^{2}g_{{1,2}}-9\,g_{{3,3}}g_{{2,3}}g_{{2,1}}+9\,g_{{1,2}}g_{{1,3}}g
_{{3,3}}-g_{{2,2}}{g_{{3,3}}}^{2}
\end{array}
$$\\
\\
And we have checked that
$$
\{H_{2}\,,\,H_{2}\}_{3}=0.
$$\\
\\
\textbf{The case N=4}
\\
\\

$ H_{2}= 2\,g_{{1,1}}g_{{2,2}}+ 6\,g_{{1,1}}g_{{3,3}}-
5\,{g_{{4,4}}}^{2 }+ 6\,g_{{4,4}}g_{{2,2}}+ 2\,g_{{4,4}}g_{{3,3}}+
32\,g_{{3,4}}g_{{3,2}}+ 2 \,g_{{2,2}}g_{{3,3}}+
32\,g_{{1,3}}g_{{2,4}}+ 32\,g_{{2,3}}g_{{2,1}}+ 64
\,{g_{{2,4}}}^{2} + 32\,g_{{2,4}}g_{{3,1}}- 5\,{g_{{1,1}}}^{2}+
2\,g_{{1,1
}}g_{{4,4}}+64\,g_{{3,4}}g_{{2,1}}+128\,g_{{1,4}}g_{{3,4}}-5\,{g_{{2,2
}}}^{2}+32\,g_{{1,2}}g_{{1,4}}-5\,{g_{{3,3}}}^{2}+64\,g_{{2,3}}g_{{1,4
}} $\\
\\
 $ H_{3}=
4\,g_{{1,3}}g_{{2,4}}g_{{4,4}}-4\,g_{{1,3}}g_{{2,4}}g_{{3
,3}}+4\,g_{{1,3}}g_{{2,4}}g_{{2,2}}-16\,{g_{{3,4}}}^{2}g_{{1,3}}-{g_{{
1,1}}}^{2}g_{{4,4}}-4\,g_{{1,3}}g_{{2,4}}g_{{1,1}}-64\,g_{{3,4}}g_{{2,
3}}g_{{2,4}}-32\,g_{{1,2}}g_{{3,4}}g_{{2,4}}
-8\,g_{{3,4}}g_{{2,1}}g_{{
4,4}}-16\,g_{{3,4}}g_{{2,3}}g_{{3,1}}+4\,g_{{2,4}}g_{{3,1}}g_{{1,1}}+8
\,g_{{3,4}}g_{{2,1}}g_{{1,1}}-8\,g_{{2,3}}g_{{2,1}}g_{{3,3}}+4\,g_{{2,
4}}g_{{3,1}}g_{{3,3}}-4\,g_{{2,4}}g_{{3,1}}g_{{2,2}}-4\,g_{{2,4}}g_{{3
,1}}g_{{4,4}}
-8\,g_{{3,4}}g_{{3,2}}g_{{4,4}}+8\,g_{{1,2}}g_{{1,4}}g_{{
4,4}}-8\,g_{{3,4}}g_{{2,1}}g_{{3,3}}-8\,g_{{2,3}}g_{{1,4}}g_{{3,3}}+8
\,g_{{2,3}}g_{{2,1}}g_{{1,1}}-16\,g_{{1,2}}g_{{2,3}}g_{{2,4}}+8\,g_{{2
,3}}g_{{1,4}}g_{{1,1}}+8\,g_{{3,4}}g_{{3,2}}g_{{2,2}}
+8\,g_{{2,3}}g_{{
1,4}}g_{{4,4}}-8\,g_{{1,2}}g_{{1,4}}g_{{2,2}}+16\,g_{{1,3}}g_{{2,1}}g_
{{1,4}}+8\,g_{{3,4}}g_{{2,1}}g_{{2,2}}+64\,g_{{2,4}}g_{{2,1}}g_{{1,4}}
+32\,g_{{2,4}}g_{{3,2}}g_{{1,4}}-8\,g_{{2,3}}g_{{1,4}}g_{{2,2}}+16\,g_
{{1,3}}{g_{{1,4}}}^{2}
+16\,g_{{2,4}}{g_{{2,1}}}^{2}+64\,g_{{2,4}}{g_{{
1,4}}}^{2}-16\,{g_{{3,4}}}^{2}g_{{3,1}}-16\,{g_{{2,3}}}^{2}g_{{2,4}}+
16\,{g_{{1,4}}}^{2}g_{{3,1}}-64\,{g_{{3,4}}}^{2}g_{{2,4}}+{g_{{2,2}}}^
{2}g_{{3,3}}-g_{{2,2}}{g_{{3,3}}}^{2}+g_{{4,4}}{g_{{3,3}}}^{2}
-g_{{1,1
}}{g_{{2,2}}}^{2}+g_{{1,1}}{g_{{4,4}}}^{2}-{g_{{4,4}}}^{2}g_{{3,3}}+16
\,g_{{2,4}}g_{{3,2}}g_{{2,1}}+{g_{{1,1}}}^{2}g_{{2,2}}
 $\\
 \\
$
 H_{4}= 49152\,{g_{{2,4}}}^{2}g_{{1,4}}g_{{3,4}}+ 3584\,
g_{{4,4}}g_{{2,2}}g_{{3,4}}g_{{1,4}}+
1536\,g_{{4,4}}g_{{1,2}}g_{{2,3}} g_{{2,4}}-
1536\,g_{{2,4}}g_{{3,2}}g_{{2,1}}g_{{1,1}}+ 8192\,g_{{1,3}}g_
{{2,4}}g_{{2,3}}g_{{1,4}}\\-
 256\,g_{{3,4}}g_{{2,1}}g_{{1,1}}g_{{4,4}}+
 512\,g_{{2,2}}g_{{3,3}}g_{{1,4}}g_{{3,4}}+4096\,g_{{1,3}}g_{{2,1}}g_{{
2,3}}g_{{2,4}}-2048\,g_{{2,4}}g_{{2,2}}g_{{3,2}}g_{{1,4}}+4096\,g_{{2,
4}}g_{{3,2}}g_{{3,4}}g_{{3,1}}+
1792\,g_{{4,4}}g_{{2,2}}g_{{2,3}}g_{{1,4}}
+1536\,g_{{3,3}}g_{{1,3}}g_{{2,1}}g_{{1,4}}-128\,g_{{1,1}}g_{{2,2}}
g_{{3,4}}g_{{3,2}}+384\,g_{{4,4}}g_{{3,3}}g_{{3,4}}g_{{3,2}}+4096\,g_{
{2,3}}g_{{2,1}}g_{{2,4}}g_{{3,1}}+
384\,g_{{2,3}}g_{{2,1}}g_{{2,2}}g_{{
3,3}}+3584\,g_{{1,1}}g_{{3,3}}g_{{3,4}}g_{{1,4}}
+384\,g_{{2,2}}g_{{3,3
}}g_{{3,4}}g_{{3,2}}+512\,g_{{1,3}}g_{{2,1}}g_{{1,4}}g_{{2,2}}-128\,g_
{{2,3}}g_{{2,1}}g_{{4,4}}g_{{3,3}}-
512\,g_{{2,2}}g_{{1,2}}g_{{2,3}}g_{
{2,4}}+384\,g_{{1,1}}g_{{3,3}}g_{{1,2}}g_{{1,4}}-1536\,g_{{3,4}}g_{{2,
3}}g_{{3,1}}g_{{4,4}}-
128\,g_{{4,4}}g_{{3,3}}g_{{1,2}}g_{{1,4}}+384\,g
_{{1,1}}g_{{2,2}}g_{{1,2}}g_{{1,4}}+
24576\,g_{{3,4}}g_{{2,3}}g_{{2,1}}
g_{{1,4}}+1408\,g_{{4,4}}g_{{2,2}}g_{{3,4}}g_{{3,2}}+256\,g_{{3,4}}g_{
{2,1}}g_{{4,4}}g_{{3,3}}+4096\,g_{{3,4}}g_{{2,3}}g_{{3,2}}g_{{1,4}}
- 512\,g_{{2,4}}g_{{2,2}}g_{{3,2}}g_{{2,1}}+
1408\,g_{{4,4}}g_{{2,2}}g_{{
1,2}}g_{{1,4}}+768\,g_{{3,4}}g_{{2,1}}g_{{2,2}}g_{{3,3}}+4096\,g_{{3,4
}}g_{{1,3}}g_{{3,2}}g_{{2,4}}+4096\,g_{{3,4}}g_{{2,3}}g_{{3,2}}g_{{2,1
}}+256\,g_{{1,1}}g_{{4,4}}g_{{2,3}}g_{{1,4}} +
1536\,g_{{2,4}}g_{{4,4}}g
_{{3,2}}g_{{2,1}}+384\,g_{{1,1}}g_{{3,3}}g_{{3,4}}g_{{3,2}}+2048\,g_{{
4,4}}g_{{3,4}}g_{{2,3}}g_{{2,4}}+1408\,g_{{2,3}}g_{{2,1}}g_{{1,1}}g_{{
3,3}}-128\,g_{{2,3}}g_{{2,1}}g_{{1,1}}g_{{4,4}}+
384\,g_{{2,3}}g_{{2,1} }g_{{4,4}}g_{{2,2}}
+4096\,g_{{1,3}}g_{{2,4}}g_{{1,2}}g_{{1,4}}+1536\,g
_{{1,1}}g_{{2,3}}g_{{3,4}}g_{{3,1}}-2048\,g_{{2,4}}g_{{2,1}}g_{{1,4}}g
_{{2,2}}-1536\,g_{{1,3}}g_{{2,1}}g_{{1,4}}g_{{4,4}}-
128\,g_{{1,1}}g_{{
4,4}}g_{{3,4}}g_{{3,2}}-1536\,g_{{1,2}}g_{{2,3}}g_{{2,4}}g_{{3,3}}+
2048\,g_{{2,4}}g_{{3,3}}g_{{2,1}}g_{{1,4}}-128\,g_{{2,2}}g_{{3,3}}g_{{
1,2}}g_{{1,4}}+24576\,g_{{1,3}}g_{{2,4}}g_{{3,4}}g_{{1,4}}+
12288\,g_{{
1,4}}g_{{3,1}}g_{{2,3}}g_{{2,4}}+8192\,g_{{3,4}}g_{{2,1}}g_{{2,4}}g_{{
3,1}}+256\,g_{{2,2}}g_{{3,3}}g_{{2,3}}g_{{1,4}}+512\,g_{{3,4}}g_{{2,3}
}g_{{3,1}}g_{{2,2}}+1792\,g_{{3,4}}g_{{2,1}}g_{{1,1}}g_{{3,3}}-
448\,g_
{{2,4}}g_{{3,1}}{g_{{1,1}}}^{2}+384\,g_{{2,3}}g_{{2,1}}g_{{1,1}}g_{{2,
2}}-
448\,g_{{2,4}}g_{{3,1}}{g_{{3,3}}}^{2}+1792\,g_{{1,1}}g_{{3,3}}g_{
{2,3}}g_{{1,4}}+
4096\,g_{{1,2}}g_{{2,3}}g_{{2,1}}g_{{1,4}}+2048\,g_{{1
,4}}g_{{3,2}}g_{{2,4}}g_{{4,4}}
-448\,g_{{2,4}}g_{{3,1}}{g_{{2,2}}}^{2}
-448\,g_{{2,4}}g_{{3,1}}{g_{{4,4}}}^{2}+512\,g_{{1,1}}g_{{4,4}}g_{{1,4
}}g_{{3,4}}-832\,{g_{{4,4}}}^{2}g_{{3,4}}g_{{3,2}}+
4096\,g_{{1,2}}g_{{
3,4}}g_{{2,1}}g_{{1,4}}+384\,g_{{1,1}}g_{{4,4}}g_{{1,2}}g_{{1,4}}+
24576\,{g_{{2,4}}}^{2}g_{{2,1}}g_{{3,4}}+
24576\,g_{{2,3}}{g_{{1,4}}}^{
2}g_{{3,4}}+1792\,g_{{3,4}}g_{{2,1}}g_{{4,4}}g_{{2,2}}+36\,{g_{{1,1}}}
^{2}g_{{4,4}}g_{{3,3}}-
256\,g_{{4,4}}g_{{3,3}}g_{{2,3}}g_{{1,4}}-180\,
{g_{{1,1}}}^{2}g_{{4,4}}g_{{2,2}}-
512\,g_{{3,3}}g_{{2,3}}g_{{3,4}}g_{{
3,1}}-1536\,g_{{1,3}}{g_{{1,4}}}^{2}g_{{4,4}}+36\,{g_{{4,4}}}^{2}g_{{2
,2}}g_{{3,3}}+36\,g_{{1,1}}{g_{{3,3}}}^{2}g_{{4,4}}+
4096\,g_{{2,4}}g_{
{1,2}}g_{{3,1}}g_{{1,4}}-180\,g_{{1,1}}{g_{{4,4}}}^{2}g_{{3,3}}+
36\,g_{{1,1}}{g_{{4,4}}}^{2}g_{{2,2}}+
36\,{g_{{1,1}}}^{2}g_{{2,2}}g_{{3,3}}+
36\,g_{{1,1}}{g_{{2,2}}}^{2}g_{{4,4}}+2048\,g_{{4,4}}g_{{1,2}}g_{{3,4}
}g_{{2,4}}+512\,{g_{{1,4}}}^{2}g_{{1,1}}g_{{3,1}}
+12288\,g_{{1,4}}g_{{ 3,2}}{g_{{3,4}}}^{2}+
768\,g_{{1,1}}g_{{2,2}}g_{{2,3}}g_{{1,4}}-2048\,g
_{{2,2}}g_{{2,3}}g_{{2,4}}g_{{3,4}}-512\,g_{{1,1}}g_{{1,3}}g_{{2,1}}g_
{{1,4}}-1536\,{g_{{3,4}}}^{2}g_{{3,1}}g_{{4,4}}-1536\,{g_{{2,3}}}^{2}g
_{{2,4}}g_{{3,3}}+ 12288\,g_{{1,3}}g_{{2,1}}g_{{2,4}}g_{{3,4}}
+8192\,g_
{{1,2}}{g_{{1,4}}}^{2}g_{{2,3}}+12288\,g_{{1,2}}{g_{{1,4}}}^{2}g_{{3,4
}}-2048\,g_{{1,4}}g_{{2,1}}g_{{2,4}}g_{{1,1}}-2048\,g_{{2,2}}g_{{1,2}}
g_{{2,4}}g_{{3,4}}+24576\,g_{{2,4}}g_{{3,4}}g_{{3,1}}g_{{1,4}}+
2048\,g
_{{1,4}}g_{{2,1}}g_{{2,4}}g_{{4,4}}+256\,g_{{3,4}}g_{{2,1}}g_{{1,1}}g_
{{2,2}}+512\,g_{{1,1}}g_{{2,2}}g_{{3,4}}g_{{1,4}}+4096\,g_{{3,4}}g_{{1
,3}}g_{{3,1}}g_{{1,4}}-448\,{g_{{1,1}}}^{2}g_{{1,3}}g_{{2,4}}+1536\,g_
{{1,1}}{g_{{3,4}}}^{2}g_{{3,1}} -
64\,{g_{{3,3}}}^{2}g_{{1,2}}g_{{1,4}}-
512\,g_{{1,1}}g_{{1,3}}{g_{{1,4}}}^{2}+512\,{g_{{3,4}}}^{2}g_{{3,1}}g_
{{2,2}}-180\,g_{{1,1}}{g_{{2,2}}}^{2}g_{{3,3}}+1536\,g_{{4,4}}{g_{{2,3
}}}^{2}g_{{2,4}}+768\,g_{{1,1}}g_{{4,4}}{g_{{2,4}}}^{2}-
512\,g_{{2,2}} {g_{{2,3}}}^{2}g_{{2,4}}+
24576\,g_{{1,4}}g_{{2,1}}{g_{{3,4}}}^{2}+768
\,g_{{1,1}}g_{{2,2}}{g_{{2,4}}}^{2}+36\,g_{{1,1}}g_{{2,2}}{g_{{3,3}}}^
{2}-180\,g_{{4,4}}g_{{2,2}}{g_{{3,3}}}^{2}+36\,g_{{4,4}}{g_{{2,2}}}^{2
}g_{{3,3}}-
448\,{g_{{4,4}}}^{2}g_{{1,3}}g_{{2,4}}+256\,g_{{1,1}}g_{{3,
3}}{g_{{2,4}}}^{2}+
256\,g_{{4,4}}g_{{2,2}}{g_{{2,4}}}^{2}-2304\,{g_{{4
,4}}}^{2}g_{{1,4}}g_{{3,4}}+12288\,{g_{{2,4}}}^{2}g_{{3,4}}g_{{3,2}}+
768\,g_{{2,2}}g_{{3,3}}{g_{{2,4}}}^{2}
+512\,{g_{{2,3}}}^{2}g_{{2,4}}g_
{{1,1}}-2048\,g_{{3,4}}g_{{2,3}}g_{{2,4}}g_{{3,3}}-
448\,{g_{{2,2}}}^{2
}g_{{1,3}}g_{{2,4}}+768\,g_{{4,4}}g_{{3,3}}{g_{{2,4}}}^{2}+512\,g_{{2,
4}}{g_{{2,1}}}^{2}g_{{3,3}}+24576\,{g_{{2,4}}}^{2}g_{{1,4}}g_{{2,3}}
+
1536\,g_{{2,4}}g_{{4,4}}{g_{{2,1}}}^{2}-512\,g_{{3,3}}{g_{{3,4}}}^{2}g
_{{3,1}}-448\,g_{{1,3}}g_{{2,4}}{g_{{3,3}}}^{2}
+512\,g_{{1,3}}{g_{{1,4
}}}^{2}g_{{2,2}}+1536\,g_{{3,3}}g_{{1,3}}{g_{{1,4}}}^{2}-512\,g_{{2,4}
}g_{{2,2}}{g_{{2,1}}}^{2}
-1408\,g_{{3,4}}g_{{2,1}}{g_{{2,2}}}^{2}-832
\,{g_{{4,4}}}^{2}g_{{1,2}}g_{{1,4}}-896\,g_{{3,4}}g_{{2,1}}{g_{{1,1}}}
^{2}+
12288\,{g_{{2,4}}}^{2}g_{{1,2}}g_{{1,4}}-1408\,{g_{{1,1}}}^{2}g_{
{2,3}}g_{{1,4}}-576\,{g_{{3,3}}}^{2}g_{{3,4}}g_{{3,2}}
-832\,{g_{{2,2}}
}^{2}g_{{3,4}}g_{{3,2}}-1408\,{g_{{2,2}}}^{2}g_{{2,3}}g_{{1,4}}-1536\,
g_{{2,4}}{g_{{2,1}}}^{2}g_{{1,1}}+
4096\,{g_{{2,4}}}^{2}g_{{1,2}}g_{{3,
2}}-2304\,{g_{{1,1}}}^{2}g_{{1,4}}g_{{3,4}}-1408\,g_{{3,4}}g_{{2,1}}{g
_{{3,3}}}^{2}
+4096\,{g_{{2,4}}}^{2}g_{{2,3}}g_{{3,2}}-896\,{g_{{3,3}}}
^{2}g_{{2,3}}g_{{1,4}}-1536\,{g_{{1,4}}}^{2}g_{{3,3}}g_{{3,1}}-
2304\,{
g_{{3,3}}}^{2}g_{{3,4}}g_{{1,4}}-512\,g_{{4,4}}g_{{1,3}}{g_{{3,4}}}^{2
}-832\,{g_{{2,2}}}^{2}g_{{1,2}}g_{{1,4}}
-896\,g_{{3,4}}g_{{2,1}}{g_{{4
,4}}}^{2}-4096\,g_{{1,2}}g_{{2,3}}{g_{{3,4}}}^{2}-64\,{g_{{1,1}}}^{2}g
_{{3,4}}g_{{3,2}}+1536\,g_{{2,2}}g_{{1,3}}{g_{{3,4}}}^{2}
-896\,{g_{{4,
4}}}^{2}g_{{2,3}}g_{{1,4}}+1536\,{g_{{1,4}}}^{2}g_{{3,1}}g_{{2,2}}
+4096\,{g_{{2,4}}}^{2}g_{{1,2}}g_{{2,1}}-2304\,{g_{{2,2}}}^{2}g_{{3,4}}
g_{{1,4}}
+8192\,{g_{{2,3}}}^{2}g_{{2,1}}g_{{1,4}}-64\,g_{{2,3}}g_{{2,1
}}{g_{{4,4}}}^{2}-
576\,{g_{{1,1}}}^{2}g_{{1,2}}g_{{1,4}}+8192\,g_{{3,4
}}{g_{{2,1}}}^{2}g_{{2,3}}
+8192\,{g_{{3,4}}}^{2}g_{{3,2}}g_{{2,1}}-512
\,{g_{{1,4}}}^{2}g_{{3,1}}g_{{4,4}}+512\,g_{{3,3}}g_{{1,3}}{g_{{3,4}}}
^{2}-832\,g_{{2,3}}g_{{2,1}}{g_{{1,1}}}^{2}-1536\,g_{{1,1}}g_{{1,3}}{g
_{{3,4}}}^{2} -4096\,{g_{{1,4}}}^{2}g_{{3,2}}g_{{2,1}}
+12288\,{g_{{2,4}
}}^{2}g_{{2,1}}g_{{2,3}}-576\,g_{{2,3}}g_{{2,1}}{g_{{2,2}}}^{2}-832\,g
_{{2,3}}g_{{2,1}}{g_{{3,3}}}^{2}+2048\,g_{{3,4}}g_{{2,3}}g_{{2,4}}g_{{
1,1}}+896\,g_{{2,2}}g_{{3,3}}g_{{1,3}}g_{{2,4}}-
128\,g_{{4,4}}g_{{3,3} }g_{{1,3}}g_{{2,4}}
+128\,g_{{4,4}}g_{{2,2}}g_{{1,3}}g_{{2,4}}+128\,g_{
{1,1}}g_{{3,3}}g_{{1,3}}g_{{2,4}}-128\,g_{{1,1}}g_{{2,2}}g_{{1,3}}g_{{
2,4}}+896\,g_{{1,1}}g_{{4,4}}g_{{1,3}}g_{{2,4}}+512\,g_{{4,4}}g_{{3,3}
}g_{{3,4}}g_{{1,4}}+
216\,g_{{1,1}}g_{{2,2}}g_{{4,4}}g_{{3,3}}+896\,g_{
{2,4}}g_{{3,1}}g_{{1,1}}g_{{2,2}}+896\,g_{{2,4}}g_{{3,1}}g_{{4,4}}g_{{
3,3}}+128\,g_{{2,4}}g_{{3,1}}g_{{4,4}}g_{{2,2}}
+128\,g_{{2,4}}g_{{3,1} }g_{{1,1}}g_{{3,3}}-
128\,g_{{2,4}}g_{{3,1}}g_{{1,1}}g_{{4,4}}- 128\,g_{
{2,4}}g_{{3,1}}g_{{2,2}}g_{{3,3}}+512\,g_{{1,2}}g_{{2,3}}g_{{2,4}}g_{{
1,1}}-60\,{g_{{1,1}}}^{3}g_{{3,3}}-52\,{g_{{1,1}}}^{3}g_{{2,2}}-52\,{g
_{{1,1}}}^{3}g_{{4,4}}+
2048\,{g_{{2,4}}}^{2}{g_{{3,1}}}^{2}-896\,{g_{{
3,3}}}^{2}{g_{{2,4}}}^{2}+150\,{g_{{4,4}}}^{2}{g_{{3,3}}}^{2}
+54\,{g_{
{4,4}}}^{2}{g_{{2,2}}}^{2}-52\,{g_{{4,4}}}^{3}g_{{3,3}}-60\,{g_{{4,4}}
}^{3}g_{{2,2}}+54\,{g_{{1,1}}}^{2}{g_{{3,3}}}^{2}+
150\,{g_{{1,1}}}^{2}
{g_{{2,2}}}^{2}-52\,g_{{1,1}}{g_{{4,4}}}^{3}+150\,{g_{{1,1}}}^{2}{g_{{
4,4}}}^{2}-896\,{g_{{1,1}}}^{2}{g_{{2,4}}}^{2}-
8192\,{g_{{3,4}}}^{3}g_
{{2,3}}+8192\,{g_{{2,3}}}^{2}{g_{{1,4}}}^{2}-4096\,{g_{{1,4}}}^{2}{g_{
{2,1}}}^{2}
-4096\,g_{{1,2}}{g_{{3,4}}}^{3}-4096\,{g_{{2,3}}}^{2}{g_{{3
,4}}}^{2}-4096\,{g_{{1,4}}}^{3}g_{{3,2}}+4096\,g_{{1,3}}{g_{{2,4}}}^{3
}+2048\,{g_{{1,3}}}^{2}{g_{{2,4}}}^{2}
-52\,g_{{2,2}}{g_{{3,3}}}^{3}+ 150\,{g_{{2,2}}}^{2}{g_{{3,3}}}^{2}
-896\,{g_{{2,2}}}^{2}{g_{{2,4}}}^{2
}-52\,{g_{{2,2}}}^{3}g_{{3,3}}-8192\,{g_{{1,4}}}^{3}g_{{2,1}}-52\,g_{{
4,4}}{g_{{3,3}}}^{3}-60\,g_{{4,4}}{g_{{2,2}}}^{3}-896\,{g_{{4,4}}}^{2}
{g_{{2,4}}}^{2}-60\,g_{{1,1}}{g_{{3,3}}}^{3}
-52\,g_{{1,1}}{g_{{2,2}}}^{3}+2048\,{g_{{3,4}}}^{2}{g_{{3,2}}}^{2}+2048\,{g_{{1,2}}}^{2}{g_{{1,4
}}}^{2}+8192\,{g_{{3,4}}}^{2}{g_{{2,1}}}^{2}+2048\,{g_{{2,3}}}^{2}{g_{
{2,1}}}^{2}+24576\,{g_{{3,4}}}^{2}{g_{{1,4}}}^{2}-4096\,{g_{{1,4}}}^{4
}+41\,{g_{{1,1}}}^{4}+41\,{g_{{4,4}}}^{4}+41\,{g_{{2,2}}}^{4}+4096\,{g
_{{2,4}}}^{3}g_{{3,1}}-4096\,{g_{{3,4}}}^{4}+41\,{g_{{3,3}}}^{4}+4096
\,{g_{{2,4}}}^{4}+512\,g_{{2,4}}g_{{3,2}}g_{{2,1}}g_{{3,3}} $
\\
\\
And we have checked the following commutativity properties:\\
\\
$$
\begin{array}{l}
\{H_{2}\,,\,H_{3}\}_{4}=0,\\
\\
\{H_{2}\,,\,H_{4}\}_{4}=0,\\
\\
\{H_{3}\,,\,H_{4}\}_{4}=0.\\
\end{array}
$$
\newpage
\begin{Large}\textbf{Appendix R} \end{Large}\\
\\
Here we present the examples of the rational Lax operators for
N=2,3,4. We use the following normalization:
$$
L^{RT}(z)=\dfrac{N}{z}\,L(z).
$$
\textbf{The operator L(z) for N=2}\\

$$
\begin{array}{l}
L_{1\,1}=(g_{1,1}-g_{2,2})/2\\
\\
L_{1\,2}=g_{1\,2}\\
\\
L_{2\,1}=g_{2,1}+z^3\,(g_{1,1}-g_{2,2})/2\\
\\
L_{2\,2}=-(g_{1,1}-g_{2,2})/2
\end{array}
$$
\textbf{The operator L(z) for N=3}\\
$$
\begin{array}{l}
L_{1\,1}={\frac
{16}{27}}\,i{z}^{3}g_{{1,3}}-4/3\,{z}^{2}g_{{2,3}}-1/3\,g_{{2,2
}}+2/3\,g_{{1,1}}-1/3\,g_{{3,3}}+2/3\,izg_{{1,2}}\\
\\
L_{1\,2}=-4/3\,{z}^{2}g_{{1,3}}-2\,izg_{{2,3}}+g_{{1,2}}\\
\\
L_{1\,3}=g_{{1,3}}\\
\\
L_{2\,1}=-{\frac {32}{81}}\,{z}^{4}g_{{1,3}}-{\frac
{8}{27}}\,i{z}^{3}g_{{2,3}} -{\frac {8}{9}}\,{z}^{2}g_{{1,2}}+
\left( -2/3\,ig_{{1,1}}+2/3\,ig_{{3 ,3}} \right) z+g_{{2,1}}\\
\\
L_{2\,2}=-{\frac
{8}{9}}\,i{z}^{3}g_{{1,3}}+2/3\,g_{{2,2}}-1/3\,g_{{1,1}}-1/3\,
g_{{3,3}}\\
\\
L_{2\,3}=2/3\,izg_{{1,3}}+g_{{2,3}}\\
\\
L_{3\,1}=-{\frac {128}{729}}\,{z}^{6}g_{{1,3}}+{\frac
{32}{81}}\,i{z}^{5}g_{{2, 3}}-{\frac {32}{81}}\,{z}^{4}g_{{1,2}}+
\left( -{\frac {16}{27}}\,ig_{ {1,1}}-{\frac
{8}{27}}\,ig_{{3,3}}+{\frac {8}{9}}\,ig_{{2,2}} \right)
{z}^{3}-4/3\,{z}^{2}g_{{2,1}}-2/3\,izg_{{3,2}}+g_{{3,1}}\\
\\
L_{3\,2}=-{\frac {32}{81}}\,i{z}^{5}g_{{1,3}}-{\frac
{32}{27}}\,{z}^{4}g_{{2,3} }+{\frac {8}{27}}\,i{z}^{3}g_{{1,2}}+
\left( -4/3\,g_{{1,1}}+4/3\,g_{{ 3,3}} \right)
{z}^{2}+2\,izg_{{2,1}}+g_{{3,2}}\\
\\
L_{3\,3}=4/3\,{z}^{2}g_{{2,3}}+2/3\,g_{{3,3}}-1/3\,g_{{1,1}}+{\frac
{8}{27}}\,i{z}^{3}g_{{1,3}}-1/3\,g_{{2,2}}-2/3\,izg_{{1,2}}
\end{array}
$$\\
\\
\textbf{The operator L(z) for N=4}\\
$$
\begin{array}{l}
L_{1\,1}=(3/4\,{z}^{4}g_{{1,4}}+2\,i{z}^{3}g_{{2,4}}+ \left(
-4\,g_{{3,4}}+3\,g_{{1,3}} \right) {z}^{2}+ \left(
4\,ig_{{1,2}}+4\,ig_{{2,3}} \right) z-
g_{{3,3}}-g_{{4,4}}+3\,g_{{1,1}}-g_{{2,2}})/4\\
\\
L_{1\,2}=i{z}^{3}g_{{1,4}}-2\,{z}^{2}g_{{2,4}}+ \left(
-2\,ig_{{3,4}}+ig_{{1,3}} \right) z+g_{{1,2}}\\
\\
L_{1\,3}=g_{{1,3}}-3/2\,{z}^{2}g_{{1,4}}-2\,izg_{{2,4}}\\
\\
L_{1\,4}=g_{{1,4}}\\
\\
L_{2\,1}={\frac
{3}{32}}\,i{z}^{5}g_{{1,4}}-1/16\,{z}^{4}g_{{2,4}}+3/8\,i{z}^{3}g_{{1,3}}+
\left( -1/4\,g_{{2,3}}-3/4\,g_{{1,2}} \right) {z}^{2}+
 \left( 1/2\,ig_{{4,4}}+1/2\,ig_{{3,3}}-ig_{{1,1}} \right)
 z+g_{{2,1}}\\
 \\
 L_{2\,2}=-1/2\,{z}^{4}g_{{1,4}}+ \left( -1/2\,g_{{1,3}}-g_{{3,4}} \right) {z}^{
2}+3/4\,g_{{2,2}}-1/4\,g_{{3,3}}-1/4\,g_{{4,4}}-1/4\,g_{{1,1}}\\
\\
L_{2\,3}=-3/4\,i{z}^{3}g_{{1,4}}-1/2\,{z}^{2}g_{{2,4}}+ \left(
1/2\,ig_{{1,3}}- 2\,ig_{{3,4}} \right) z+g_{{2,3}}\\
\\
L_{2\,4}=1/2\,izg_{{1,4}}+g_{{2,4}}\\
\\
\L_{3\,1}=-{\frac
{3}{64}}\,{z}^{6}g_{{1,4}}+1/16\,i{z}^{5}g_{{2,4}}+ \left( -1/
16\,g_{{3,4}}-3/16\,g_{{1,3}} \right) {z}^{4}+ \left(
1/4\,ig_{{2,3}}- 3/8\,ig_{{1,2}} \right) {z}^{3}+ \left(
3/4\,g_{{1,1}}-1/4\,g_{{3,3}}- 1/2\,g_{{2,2}} \right) {z}^{2}\\
\\+ \left( -1/2\,ig_{{3,2}}-ig_{{2,1}}
 \right) z+g_{{3,1}}\\
 \\
 L_{3\,2}=-1/4\,i{z}^{5}g_{{1,4}}-1/2\,{z}^{4}g_{{2,4}}+ \left( -1/2\,ig_{{3,4}}
-1/4\,ig_{{1,3}} \right) {z}^{3}+ \left( -1/4\,g_{{1,2}}-g_{{2,3}}
 \right) {z}^{2}+ \left( -ig_{{1,1}}+ig_{{4,4}} \right)
 z+g_{{3,2}}\\
 \\
 \end{array}
 $$

 $$
 \begin{array}{l}
 L_{3\,3}=3/8\,{z}^{4}g_{{1,4}}-i{z}^{3}g_{{2,4}}+ \left( -1/4\,g_{{1,3}}+1/2\,g
_{{3,4}} \right)
{z}^{2}-1/2\,izg_{{1,2}}-1/4\,g_{{4,4}}-1/4\,g_{{1,1}
}-1/4\,g_{{2,2}}+3/4\,g_{{3,3}}\\
\\
L_{3\,4}=izg_{{2,4}}-1/4\,{z}^{2}g_{{1,4}}+g_{{3,4}}\\
\\
L_{4\,1}=-{\frac
{3}{256}}\,{z}^{8}g_{{1,4}}+1/16\,i{z}^{7}g_{{2,4}}+ \left( -{
\frac {3}{64}}\,g_{{1,3}}+{\frac {3}{32}}\,g_{{3,4}} \right)
{z}^{6}+
 \left( 1/4\,ig_{{2,3}}-{\frac {3}{32}}\,ig_{{1,2}} \right)
 {z}^{5}+\\
 \\
 \left( -1/2\,g_{{2,2}}+3/16\,g_{{1,1}}-1/16\,g_{{4,4}}+3/8\,g_{{3,3}}
 \right) {z}^{4}+ \left( -ig_{{2,1}}+3/4\,ig_{{3,2}} \right) {z}^{3}\\
 \\+
 \left( -1/4\,g_{{4,3}}-3/2\,g_{{3,1}} \right) {z}^{2}-1/2\,izg_{{4,2}
}+g_{{4,1}}\\
\\
L_{4\,2}=-1/16\,i{z}^{7}g_{{1,4}}-3/8\,{z}^{6}g_{{2,4}}+ \left(
-1/16\,ig_{{1,3 }}+5/8\,ig_{{3,4}} \right) {z}^{5}+ \left(
-1/16\,g_{{1,2}}-1/2\,g_{{2 ,3}} \right) {z}^{4}+\\
\\
 \left( -1/2\,ig_{{1,1}}+ig_{{3,3}}-1/2\,ig_{{4,4 }} \right)
{z}^{3}+ \left( -1/2\,g_{{3,2}}-2\,g_{{2,1}} \right) {z}^{2 }+
\left(
-ig_{{4,3}}+2\,ig_{{3,1}} \right) z+g_{{4,2}}\\
\\
L_{4\,3}={\frac {3}{32}}\,{z}^{6}g_{{1,4}}-5/8\,i{z}^{5}g_{{2,4}}+
\left( -1/16 \,g_{{1,3}}-5/4\,g_{{3,4}} \right)
{z}^{4}+1/2\,i{z}^{3}g_{{2,3}}+
 \left( -g_{{2,2}}+3/2\,g_{{4,4}}-g_{{1,1}}+1/2\,g_{{3,3}} \right) {z}
^{2}\\
\\
+ \left( 2\,ig_{{2,1}}+2\,ig_{{3,2}} \right) z+g_{{4,3}}\\
\\
L_{4\,4}=-1/16\,{z}^{4}g_{{1,4}}+1/2\,i{z}^{3}g_{{2,4}}+3/2\,{z}^{2}g_{{3,4}}+
 \left( -ig_{{2,3}}-1/2\,ig_{{1,2}} \right) z-1/4\,g_{{3,3}}-1/4\,g_{{
2,2}}+3/4\,g_{{4,4}}-1/4\,g_{{1,1}}
\end{array}
$$
\\
\\
\textbf{The hamiltonians and highest integrals of motion}\\
\\
On calculating the integrals of motion we use the standard
definition:
$$
H_{k}=\dfrac{1}{2\,k\,\pi\,i}\oint\,\dfrac{{\rm tr}
\left(L^{RT}(z)\right)^{k}}{z}\,dz.
$$
The integral $H_{2}$ is usually identified with the hamiltonian.
We check the commutativity with respect to the standard Poisson
bracket on the algebra gl(N,$\mathbb{C}$):

$$
\{A\,,\,B\}_{N}=\sum_{i\,j\,k\,=1}^{N}\,\left(\dfrac{\partial\,A}{\partial\,g_{i,j}}\,\dfrac{\partial\,B}{\partial\,g_{j,k}}\,-
\dfrac{\partial\,B}{\partial\,g_{i,j}}\,\dfrac{\partial\,A}{\partial\,g_{j,k}}\right)\,g_{i\,k}
$$\\
\textbf{The case N=2}\\
$$
H_{2}=2\,g_{1,2}\,(g_{1,1}-g_{2,2})
$$
\textbf{The case N=3}\\
\\
$
H_{2}=-{g_{{1,2}}}^{2}+3\,g_{{3,3}}g_{{2,3}}-3\,g_{{2,3}}g_{{1,1}}-3\,g_{{2,
1}}g_{{1,3}} $\\
\\
 $
H_{3}=3\,g_{{3,3}}g_{{2,3}}g_{{1,2}}-3\,g_{{2,3}}g_{{2,2}}g_{{1,2}}+2/3\,{g_
{{1,2}}}^{3}-2\,{g_{{3,3}}}^{2}g_{{1,3}}+{g_{{2,2}}}^{2}g_{{1,3}}-9\,{
g_{{2,3}}}^{2}g_{{2,1}}+5\,g_{{3,3}}g_{{1,1}}g_{{1,3}}-g_{{2,2}}g_{{1,
1}}g_{{1,3}}
+3\,g_{{1,3}}g_{{2,1}}g_{{1,2}}-2\,{g_{{1,1}}}^{2}g_{{1,3}
}-g_{{3,3}}g_{{2,2}}g_{{1,3}} $
\\
\\
And we check the following condition for integrability:
$$
\{H_{2}\,,\,H_{3}\}_{3}=0
$$
\textbf{The case N=4}
\\
\\
$
H_{2}=-4\,g_{{1,2}}g_{{2,3}}-8\,g_{{1,4}}g_{{3,1}}-4\,{g_{{2,3}}}^{2}+6\,g_{
{1,1}}g_{{1,3}}-4\,g_{{2,2}}g_{{3,4}}-2\,g_{{2,2}}g_{{1,3}}-12\,g_{{1,
1}}g_{{3,4}}-2\,g_{{4,4}}g_{{1,3}}-2\,g_{{3,3}}g_{{1,3}}+
12\,g_{{4,4}}
g_{{3,4}}+4\,g_{{3,3}}g_{{3,4}}-3\,{g_{{1,2}}}^{2}-8\,g_{{2,4}}g_{{3,2
}}-16\,g_{{2,4}}g_{{2,1}} $
\\
\\
$
H_{3}=-8\,g_{{3,4}}g_{{2,1}}g_{{1,3}}-4\,g_{{1,4}}g_{{3,1}}g_{{1,2}}-4\,g_{{
4,4}}g_{{3,4}}g_{{1,2}}+4\,g_{{2,4}}g_{{3,2}}g_{{1,2}}-2\,g_{{2,3}}g_{
{1,3}}g_{{3,3}}-2\,g_{{1,4}}g_{{3,2}}g_{{2,2}}+4\,g_{{3,3}}g_{{2,3}}g_
{{3,4}}\\
+16\,g_{{2,4}}g_{{3,1}}g_{{3,4}}-{g_{{1,2}}}^{3}-2\,g_{{2,3}}g_
{{1,3}}g_{{4,4}}-2\,{g_{{1,2}}}^{2}g_{{2,3}}-2\,g_{{1,4}}g_{{3,2}}g_{{
4,4}}+3\,g_{{1,1}}g_{{1,2}}g_{{1,3}}-8\,g_{{1,3}}g_{{3,1}}g_{{2,4}}-8
\,g_{{1,4}}g_{{2,1}}g_{{4,4}}\\
-4\,g_{{4,4}}g_{{2,4}}g_{{2,2}}-2\,g_{{2,
3}}g_{{1,3}}g_{{2,2}}-g_{{2,2}}g_{{1,3}}g_{{1,2}}+4\,g_{{2,2}}g_{{2,3}
}g_{{3,4}}+6\,g_{{2,3}}g_{{1,3}}g_{{1,1}}+4\,{g_{{1,1}}}^{2}g_{{2,4}}+
4\,{g_{{4,4}}}^{2}g_{{2,4}}+16\,{g_{{3,4}}}^{2}g_{{2,1}}\\
+8\,{g_{{3,4}}
}^{2}g_{{3,2}}+4\,g_{{4,4}}g_{{2,4}}g_{{3,3}}-4\,g_{{1,1}}g_{{2,4}}g_{
{3,3}}-8\,g_{{1,1}}g_{{2,4}}g_{{4,4}}-4\,g_{{1,1}}g_{{3,4}}g_{{2,3}}-2
\,g_{{1,4}}g_{{3,2}}g_{{3,3}}-g_{{3,3}}g_{{1,3}}g_{{1,2}}+\\
4\,g_{{1,1}}
g_{{2,4}}g_{{2,2}}+8\,g_{{1,4}}g_{{2,1}}g_{{1,1}}+4\,g_{{2,2}}g_{{3,4}
}g_{{1,2}}+6\,g_{{1,4}}g_{{3,2}}g_{{1,1}}-g_{{4,4}}g_{{1,3}}g_{{1,2}}-
4\,g_{{4,4}}g_{{2,3}}g_{{3,4}} $\\
\\
$
H_{4}=-24\,g_{{2,2}}g_{{4,4}}g_{{1,2}}g_{{2,4}}+16\,g_{{1,2}}g_{{2,1}}
g_{{3,
4}}g_{{1,3}}-16\,g_{{2,2}}g_{{4,4}}g_{{2,4}}g_{{2,3}}-20\,g_{{2,2}}g_{
{4,4}}g_{{1,1}}g_{{1,4}}+8\,g_{{1,4}}g_{{3,2}}g_{{2,3}}g_{{4,4}}-
4\,g_{{2,2}}g_{{3,3}}g_{{1,1}}g_{{1,4}}+16\,g_{{1,3}}g_{{3,1}}g_{{3,3}}
g_{{
1,4}}+64\,g_{{1,4}}g_{{3,2}}g_{{2,4}}g_{{3,1}}+8\,g_{{1,1}}g_{{2,2}}g_
{{1,2}}g_{{2,4}}-24\,g_{{3,3}}g_{{4,4}}g_{{3,4}}g_{{1,3}}-48\,g_{{1,3}
}g_{{3,2}}g_{{2,4}}g_{{1,1}}+16\,g_{{1,3}}g_{{3,2}}g_{{2,4}}g_{{4,4}}+
16\,g_{{1,4}}g_{{2,3}}g_{{2,1}}g_{{3,3}}-16\,g_{{1,1}}g_{{4,4}}g_{{1,2
}}g_{{2,4}}-64\,g_{{4,4}}g_{{3,4}}g_{{1,4}}g_{{3,1}}-48\,g_{{4,4}}g_{{
3,4}}g_{{1,2}}g_{{2,3}}-16\,g_{{1,4}}g_{{2,3}}g_{{2,1}}g_{{1,1}}
-16\,g_{{1,4}}g_{{2,3}}g_{{2,1}}g_{{4,4}}-256\,g_{{4,4}}g_{{3,4}}g_{{2,4}}g_
{{2,1}}+16\,g_{{1,3}}g_{{3,2}}g_{{2,4}}g_{{2,2}}+16\,g_{{1,4}}g_{{2,3}
}g_{{2,1}}g_{{2,2}}+12\,g_{{4,4}}g_{{1,3}}g_{{1,2}}g_{{2,3}}+80\,g_{{4
,4}}g_{{1,3}}g_{{2,4}}g_{{2,1}}+16\,g_{{1,1}}g_{{3,3}}g_{{2,4}}g_{{2,3
}}+8\,g_{{1,1}}g_{{3,3}}g_{{1,2}}g_{{2,4}}+16\,g_{{1,2}}g_{{2,1}}g_{{4
,4}}g_{{1,4}}-32\,g_{{3,4}}g_{{1,2}}g_{{2,4}}g_{{3,1}}+32\,g_{{1,3}}g_
{{3,1}}g_{{2,4}}g_{{2,3}}+16\,g_{{1,3}}g_{{3,1}}g_{{1,2}}g_{{2,4}}-48
\,g_{{1,3}}g_{{3,1}}g_{{1,1}}g_{{1,4}}+16\,g_{{1,3}}g_{{3,1}}g_{{4,4}}
g_{{1,4}}-64\,g_{{3,3}}g_{{3,4}}g_{{2,4}}g_{{2,1}}+12\,g_{{3,3}}g_{{1,
3}}g_{{1,2}}g_{{2,3}}+16\,g_{{3,3}}g_{{1,3}}g_{{2,4}}g_{{2,1}}+16\,g_{
{1,3}}g_{{3,1}}g_{{2,2}}g_{{1,4}}+16\,g_{{2,2}}g_{{1,3}}g_{{2,4}}g_{{2
,1}}-8\,g_{{2,2}}g_{{4,4}}g_{{3,4}}g_{{1,3}}+8\,g_{{1,4}}g_{{3,2}}g_{{
2,3}}g_{{3,3}}-16\,g_{{1,2}}g_{{2,1}}g_{{1,1}}g_{{1,4}}-8\,g_{{2,2}}g_
{{3,3}}g_{{3,4}}g_{{1,3}}+4\,g_{{1,4}}g_{{3,2}}g_{{1,2}}g_{{3,3}}+4\,g
_{{1,4}}g_{{3,2}}g_{{1,2}}g_{{4,4}}+56\,g_{{1,1}}g_{{4,4}}g_{{3,4}}g_{
{1,3}}+56\,g_{{1,1}}g_{{3,3}}g_{{3,4}}g_{{1,3}}+8\,g_{{1,1}}g_{{2,2}}g
_{{3,4}}g_{{1,3}}+8\,g_{{3,3}}g_{{4,4}}g_{{1,2}}g_{{2,4}}-12\,{g_{{4,4
}}}^{2}g_{{3,4}}g_{{1,3}}+48\,g_{{1,1}}g_{{3,4}}g_{{1,2}}g_{{2,3}}+256
\,g_{{1,1}}g_{{3,4}}g_{{2,4}}g_{{2,1}}+64\,g_{{2,2}}g_{{3,4}}g_{{2,4}}
g_{{2,1}}+20\,g_{{3,3}}g_{{4,4}}g_{{1,1}}g_{{1,4}}+4\,g_{{3,3}}g_{{4,4
}}g_{{2,2}}g_{{1,4}}+32\,g_{{1,2}}g_{{2,1}}g_{{2,4}}g_{{2,3}}+16\,g_{{
2,4}}g_{{3,2}}g_{{2,2}}g_{{3,4}}-48\,g_{{2,4}}g_{{3,2}}g_{{3,3}}g_{{3,
4}}-112\,g_{{2,4}}g_{{3,2}}g_{{4,4}}g_{{3,4}}+144\,g_{{2,4}}g_{{3,2}}g
_{{1,1}}g_{{3,4}}+12\,g_{{2,2}}g_{{1,3}}g_{{1,2}}g_{{2,3}}+64\,g_{{1,4
}}g_{{3,2}}g_{{2,1}}g_{{3,4}}-8\,{g_{{1,1}}}^{2}g_{{2,4}}g_{{2,3}}+32
\,g_{{2,4}}{g_{{2,3}}}^{2}g_{{2,1}}-32\,g_{{3,3}}{g_{{2,4}}}^{2}g_{{3,
1}}+40\,{g_{{1,2}}}^{2}g_{{2,1}}g_{{2,4}}-32\,g_{{4,4}}{g_{{2,4}}}^{2}
g_{{3,1}}+10\,{g_{{1,1}}}^{2}g_{{2,2}}g_{{1,4}}+8\,{g_{{2,3}}}^{2}g_{{
2,2}}g_{{3,4}}-8\,{g_{{2,3}}}^{2}g_{{3,3}}g_{{3,4}}-56\,{g_{{2,3}}}^{2
}g_{{4,4}}g_{{3,4}}-2\,{g_{{1,1}}}^{2}g_{{3,3}}g_{{1,4}}+64\,g_{{1,1}}
g_{{3,4}}g_{{1,4}}g_{{3,1}}+28\,{g_{{1,2}}}^{2}g_{{1,4}}g_{{3,1}}-56\,
g_{{2,2}}g_{{4,4}}{g_{{3,4}}}^{2}+56\,{g_{{2,3}}}^{2}g_{{1,1}}g_{{3,4}
}+4\,{g_{{2,2}}}^{2}g_{{3,4}}g_{{1,3}}-21\,{g_{{1,2}}}^{2}g_{{1,1}}g_{
{1,3}}-104\,g_{{1,1}}g_{{4,4}}{g_{{3,4}}}^{2}-8\,g_{{2,2}}g_{{3,3}}{g_
{{3,4}}}^{2}+2\,{g_{{1,2}}}^{2}g_{{2,2}}g_{{3,4}}+7\,{g_{{1,2}}}^{2}g_
{{2,2}}g_{{1,3}}-10\,{g_{{1,2}}}^{2}g_{{3,3}}g_{{3,4}}+7\,{g_{{1,2}}}^
{2}g_{{3,3}}g_{{1,3}}+40\,g_{{3,3}}g_{{4,4}}{g_{{3,4}}}^{2}-26\,{g_{{1
,1}}}^{2}g_{{4,4}}g_{{1,4}}-12\,{g_{{3,3}}}^{2}g_{{3,4}}g_{{1,3}}+128
\,g_{{2,4}}g_{{2,1}}g_{{1,4}}g_{{3,1}}-36\,g_{{1,1}}g_{{1,3}}g_{{1,2}}
g_{{2,3}}-112\,g_{{1,1}}g_{{1,3}}g_{{2,4}}g_{{2,1}}-2\,{g_{{2,2}}}^{2}
g_{{3,3}}g_{{1,4}}+6\,{g_{{2,2}}}^{2}g_{{4,4}}g_{{1,4}}-64\,g_{{1,3}}g
_{{3,1}}{g_{{3,4}}}^{2}-12\,g_{{1,1}}g_{{3,3}}{g_{{1,3}}}^{2}+64\,{g_{
{2,4}}}^{2}g_{{3,2}}g_{{2,1}}+64\,g_{{1,4}}{g_{{2,1}}}^{2}g_{{3,4}}+12
\,g_{{1,3}}{g_{{2,3}}}^{2}g_{{2,2}}-36\,g_{{1,3}}{g_{{2,3}}}^{2}g_{{1,
1}}-60\,{g_{{1,1}}}^{2}g_{{3,4}}g_{{1,3}}+32\,g_{{2,3}}g_{{3,2}}{g_{{3
,4}}}^{2}-12\,g_{{1,1}}g_{{4,4}}{g_{{1,3}}}^{2}+4\,g_{{2,2}}g_{{4,4}}{
g_{{1,3}}}^{2}-12\,g_{{1,1}}g_{{2,2}}{g_{{1,3}}}^{2}+4\,g_{{1,4}}g_{{3
,2}}g_{{1,2}}g_{{2,2}}+32\,g_{{1,2}}g_{{2,3}}g_{{1,4}}g_{{3,1}}+4\,g_{
{2,2}}g_{{3,3}}{g_{{1,3}}}^{2}-56\,g_{{1,1}}g_{{3,3}}{g_{{3,4}}}^{2}+4
\,g_{{3,3}}g_{{4,4}}{g_{{1,3}}}^{2}+40\,g_{{1,1}}g_{{2,2}}{g_{{3,4}}}^
{2}+30\,{g_{{1,2}}}^{2}g_{{1,1}}g_{{3,4}}-8\,{g_{{3,3}}}^{2}g_{{1,2}}g
_{{2,4}}+8\,{g_{{4,4}}}^{2}g_{{2,4}}g_{{2,3}}-12\,g_{{1,4}}g_{{3,2}}g_
{{1,2}}g_{{1,1}}-6\,{g_{{3,3}}}^{2}g_{{1,1}}g_{{1,4}}-8\,{g_{{3,3}}}^{
2}g_{{2,4}}g_{{2,3}}+2\,{g_{{3,3}}}^{2}g_{{2,2}}g_{{1,4}}+16\,g_{{1,3}
}g_{{3,2}}g_{{2,4}}g_{{3,3}}-2\,{g_{{3,3}}}^{2}g_{{4,4}}g_{{1,4}}+2\,{
g_{{4,4}}}^{2}g_{{2,2}}g_{{1,4}}+12\,g_{{1,3}}{g_{{2,3}}}^{2}g_{{3,3}}
-10\,{g_{{4,4}}}^{2}g_{{3,3}}g_{{1,4}}+16\,{g_{{4,4}}}^{2}g_{{1,2}}g_{
{2,4}}+26\,{g_{{4,4}}}^{2}g_{{1,1}}g_{{1,4}}+8\,g_{{1,4}}g_{{3,2}}g_{{
2,3}}g_{{2,2}}+16\,g_{{2,4}}{g_{{2,3}}}^{2}g_{{3,2}}-24\,g_{{1,4}}g_{{
3,2}}g_{{2,3}}g_{{1,1}}+32\,g_{{1,1}}{g_{{2,4}}}^{2}g_{{3,1}}+32\,g_{{
1,3}}g_{{2,3}}g_{{2,1}}g_{{3,4}}+32\,g_{{2,2}}{g_{{2,4}}}^{2}g_{{3,1}}
+32\,g_{{1,2}}g_{{2,1}}{g_{{3,4}}}^{2}+12\,{g_{{1,2}}}^{2}g_{{2,4}}g_{
{3,2}}-22\,{g_{{1,2}}}^{2}g_{{4,4}}g_{{3,4}}+7\,{g_{{1,2}}}^{2}g_{{4,4
}}g_{{1,3}}+12\,g_{{1,3}}{g_{{2,3}}}^{2}g_{{4,4}}+48\,g_{{1,2}}{g_{{3,
4}}}^{2}g_{{3,2}}+16\,g_{{1,4}}{g_{{2,3}}}^{2}g_{{3,1}}+8\,{g_{{2,2}}}
^{2}g_{{2,4}}g_{{2,3}}+8\,{g_{{2,2}}}^{2}g_{{1,2}}g_{{2,4}}+2\,{g_{{2,
2}}}^{2}g_{{1,1}}g_{{1,4}}+4\,{g_{{2,3}}}^{4}+{\frac{21}{4}}\,{g_{{1,
2}}}^{4}+8\,g_{{1,2}}{g_{{2,3}}}^{3}-2\,{g_{{2,2}}}^{3}g_{{1,4}}+12\,{
g_{{3,3}}}^{2}{g_{{3,4}}}^{2}+60\,{g_{{4,4}}}^{2}{g_{{3,4}}}^{2}+14\,{
g_{{1,2}}}^{3}g_{{2,3}}+12\,{g_{{2,2}}}^{2}{g_{{3,4}}}^{2}+64\,{g_{{3,
4}}}^{3}g_{{3,1}}+64\,{g_{{2,4}}}^{2}{g_{{2,1}}}^{2}+6\,{g_{{1,1}}}^{3
}g_{{1,4}}+2\,{g_{{3,3}}}^{2}{g_{{1,3}}}^{2}+2\,{g_{{4,4}}}^{2}{g_{{1,
3}}}^{2}+2\,{g_{{2,2}}}^{2}{g_{{1,3}}}^{2}+60\,{g_{{1,1}}}^{2}{g_{{3,4
}}}^{2}+32\,{g_{{1,4}}}^{2}{g_{{3,1}}}^{2}+32\,{g_{{2,4}}}^{2}{g_{{3,2
}}}^{2}+2\,{g_{{3,3}}}^{3}g_{{1,4}}+18\,{g_{{1,2}}}^{2}{g_{{2,3}}}^{2}
+18\,{g_{{1,1}}}^{2}{g_{{1,3}}}^{2}-6\,{g_{{4,4}}}^{3}g_{{1,4}}
$\\
\\
And we check the following integrability conditions: $$
\{H_{2}\,,\,H_{3}\}_{4}=0,\ \ \ \{H_{2}\,,\,H_{4}\}_{4}=0\ \ \
\{H_{3}\,,\,H_{4}\}_{4}=0$$

\end{small}


\begin{thebibliography}{9}
\bibitem{Smir}
A.Smirnov, Two-Bodies systems from SL(2,$\mathbb{C}$)-tops, arXiv:
math.ds/0711.2432v1 (2007)
\bibitem{Arn}
V.I.Arnold, Mathematical Methods of Classical Mechanics, Springer,
1978 ;
\bibitem{Olsh}
M.Olshanetsky, A.Perelomov, Classical integrable
finite-dimensional systems ralated to Lie algebras, Phys.Rep 71C
313-400 (1981)

\bibitem{Ves}
M-P. Grosset and A.P. Veselov, Lame equation, quantum top and
elliptic Bernoulli Polynomials, arXiv: math-ph/0508068v2
\bibitem{LOZ}
 A.Levin, M.Olshanetsky, A.Zotov, Hitchin Systems - Symplectic
Hecke Correspondence and Two-dimensional Version. Comm.Math.Phys.
236 93-133 (2003);
\bibitem{Hit}
N.Hitchin, Stable bundles and integrable systems, Doke Math. Jour.
54), 91-114 (1987);
\bibitem{AHZ}
A.Antonov, K.Hasegawa, A.Zabrodin, On trigonometric intertwining
vectors and non-dynamical R-matrix for the Ruijsenaars model,
arXiv:hep-th/9704074 (1997);
\bibitem{GorskyZabrodin}
A.Gorsky, A.Zabrodin, Degenerations of Sklyanin algebra and
Askey-Wilson polynomials, arXiv:hep-th/9303026 (1993).


\end{thebibliography}
\end{document}